\documentclass{svproc}
\usepackage{cite}
\usepackage{amsmath,amssymb,amsfonts}
\usepackage{algorithmic}
\usepackage{graphicx}
\usepackage{textcomp}
\usepackage{xcolor}
\def\BibTeX{{\rm B\kern-.05em{\sc i\kern-.025em b}\kern-.08em
		T\kern-.1667em\lower.7ex\hbox{E}\kern-.125emX}}
	
\usepackage{url}

\begin{document}
\mainmatter              
\title{Seeking Optimum System Settings for Physical Activity Recognition on Smartwatches}
\titlerunning{Physical Activity Recognition on Smartwatches}  

\author{Muhammad Ahmad\inst{1,2,*} \and Adil Khan\inst{1} \and Manuel Mazzara\inst{1} \and
Salvatore Distefano\inst{2}}

\authorrunning{M. Ahmad et al.} 

\institute{Innopolis University, Innopolis, Kazan, Russia
\and
University of Messina, Messina, Italy \\
\email{mahmad00@gmail.com}
}

\maketitle              

\begin{abstract}
Physical activity recognition using wearable devices can provide valued information regarding an individual's degree of functional ability and lifestyle. Smartphone-based physical activity recognition is a well-studied area. However, research on smartwatch-based physical activity recognition, on the other hand, is still in its infancy. Through a large-scale exploratory study, this work aims to investigate the smartwatch-based physical activity recognition domain. A detailed analysis of various feature banks and classification methods are carried out to find the optimum system settings for the best performance of any smartwatch-based physical activity recognition system for both personal and impersonal models in real life scenarios. To further validate our hypothesis for both personal and impersonal models, we tested single subject out cross validation process for smartwatch-based physical activity recognition.

\keywords{Smartwatch, Accelerometer, Magnetometer, gyroscope, Machine Learning, Physical Activity Recognition, Health Care Services.}
\end{abstract}
\section{Introduction}
\label{sec1}

Last few years have witnessed a massive growth in consumer interest in smart devices [1]. These include smartphones, smart TVs, and recently smartwatches. Because of their increased penetration, smartwatches have claimed a handsome market share. A typical smartwatch contains a heart rate monitor, GPS, thermometer, camera and accelerometer. Thus, it can provide a variety of services. These services include temperature and pulse measurements and the number of calories consumed when performing a physical activity, such as walking, running, cycling and so forth  [2].

Recently, smartwatches have emerged in our daily lives and like their smartphone counterparts, smartwatches do contain a gyroscope, magnetometer and accelerometer sensors to support similar capabilities and applications. These applications include health applications i.e. calories consumption that require sensor based physical activity recognition. A number of calories consumed during physical activities vary from one activity to another. Therefore, the first thing to do when estimating the number of calories consumed, is to first recognize which activity have been performed [3].

This work explores the idea of physical activity recognition on smartwatches using the embedded gyroscope, magnetometer and accelerometer sensors. The gyroscope, magnetometer and accelerometer sensors are ideal for physical activity recognition. We further discussed the detailed analysis of various feature banks and classification methods to seek the best settings for optimum performance on two different types of models such as impersonal and personal model over the smartwatch. Personal models referred as the classifier is built using the data only form one specific user whereas the impersonal model referred as the classifier is built using the data form every user except the one under study. To summarize, the main contributions of this work is towards answering the following questions: 1):  Existing smartphone-based physical activity recognition studies have a number of features that were utilized for smartphone-based physical activity recognition; can they be used for smartwatch-based physical activity recognition as well. 2): Which of the most commonly used classifiers from the reference field provides the best recognition accuracy. 3): Does features permutation help to improve the accuracy. 4): Does features normalization help to improve the performance of the classifier. 5): Which model (Personal or Impersonal) works better for smartwatch-based physical activity recognition. 6): Does changing the window size affect the recognition accuracy.

Physical activity recognition in general using motion sensors especially using smartphones has been studied in recent years [4-6] and currently being studied extensively [7-8] but there are a few studies on physical activity recognition using smartwatches [9-18] in which the authors studied the role of the smartwatch in physical activity recognition system for health recommendation purposes.

For instance,  Guiry et al. [9] proposed physical activity recognition system to recognize nine different activities using five different classifiers. These activities include walking, standing, cycling, running, stair ascent, stair descent, elevator ascent and descent. However, the authors studied smartphone and smartwatch based activity recognition separately and did not combine the data from both devices. Furthermore, for real-time data collection, the author's used magnetometer, accelerometer and gyroscope sensors for smartphone and accelerometer for a smartwatch. Trost et al. [10] used sensors on hip and wrist to detect physical activities using logistic regression based classification method. In [10], Trost et al. just focused on showing the potential of using the wrist position for physical activity recognition for health recommendation system. Chernbumroong et al. [11] used a single wrist-worn accelerometer to detect five different physical activities for recommendation system. These activities include standing, sitting, lying, running and walking. However, Da-Silva et al.  [12] used the same sensor to detect the eight different physical activities for recommendation system. In addition to the activities identified in [11] Da-Silva et al. include the activity of working on a computer too. The works [13] and [14] detect the eating activity using a Hidden Markov Model (HMM) with a wrist-worn accelerometer and gyroscope sensors. Ramos-Garcia identifies the eating activity by dividing it into further sub-activities in [13] which includes resting, drinking, eating, using utensil and others. Meanwhile, Dong et al. [14] proposed to differentiate eating periods from non-eating periods to identify the eating and not-eating activity. Sen et al. [15] used both accelerometer and gyroscope sensor data to recognize the eating and non-eating activities. Furthermore, the authors used smartwatch camera to take images of the food being consumed to analyze what a user is eating.

School et al. [16] presented a feasibility study to detect and identify the smoking activity using accelerometer sensor where School et al. only reported a user-specific accuracy for smoking activity. Whereas, A. Parade et al. [17] used both accelerometer, magnetometer and gyroscope sensors to recognize smoking puffs where they just try to distinguished smoking activity from other activities. A similar work has been reported in [18] where the the authors used support vector machine (SVM) to recognize six different physical activities. These activities include standing, walking, writing, jacks, smoking and jogging. Kim et al. [19] proposed the collaborative classification approach for recognizing daily activities using smartwatches. The authors exploit a single off-the-shelf smartwatch to recognize five daily activities; namely eating, watching TV, sleeping, vacuuming and showering. For fair comparisons Kim et al. suggested using single and multiple sensors-based approaches to compare with their proposed approach with the overall accuracy of 91.5\% with the improved recall rate up to 21.5\% for each activity. Weiss et al. showed that accelerometers and gyroscopes sensors that sense user's movements and can help identify the user activity [20]. Weiss et al. propose to compare the smartphone and smartwatch-based user activity recognition in which smartwatch have the ability to recognize hand-based activities namely eating which is not possible through a smartphone. They showed that the smartwatch is able to recognize the drinking activity with 93.3\% accuracy while smartphones achieve only 77.3\%. 

Nurwanto et al. [21] proposed a daily life light sport exercise activities. These activities include squat jump, push up, sit up and walking. Nurwanto et al. focuses only on accelerometer-based signal captured by left-handed persons and the obtained signals were processed through sliding window approach with the k-nearest neighbor method and dynamic time warping as the main classifier. Furthermore, Nurwanto et al. proved that k-NN together with dynamic time warping method is an efficient method for daily exercise recognition with the accuracy of 76.67\% for push up 96.69\% for squat jump and 80\% for sit up activity. They intentionally exclude the walking activity for their experimental process because of the random patterns. Al-Naffakh et al. [22] proposed the usefulness of using smartwatch motion sensor namely gyroscope and accelerometer to perform user activity recognition. Al-Naffakh et al. proved the natures of the signals captured are sufficiently discriminative to be useful in performing activity recognition.

Different activities can be recognized using motion sensors at the wrist position i.e. drinking, eating, smoking, walking, walking down stairs, walking up stairs, running, jogging, elevator up, elevator down, doing regular exercises, writing etc. However there is no study that discusses these activities all together with the recommendations of optimum system settings. Therefore, to extend the existing works, we evaluate the combination of different feature banks and classification techniques for the reliable recognition of various activities. We also evaluate three different built-in motion sensors. Moreover, unlike the current trends, we also evaluate the effect of window size on each activity performed by the individual user and its impact on the recognition performance of simple and complex activities. Although the effect of window size on physical activity recognition has been studied by Huynh et al. in [23] but the windowing effect on complex activities is yet not fully explored. Moreover, unlike the recent works, we explore the use of both \textit{Personal} and \textit{Impersonal} models individually for each activity as explained; 1): Personal model means the classifier is trained using the data only from one specific individual whereas, the Impersonal model means the classifier is built using the data from every user except the one under study. 2): For both models, one subject cross-validation process is adopted. 3): Finally, our study suggests the optimum system settings for any activity recognition system based on smartwatches.

This work analyzes the effects of combining gyroscope, magnetometer and accelerometer sensors data for physical activity recognition for different size of windows varying from one second to 12 seconds long window. We further analyze the effect of various sampling rates on the recognition performance in different scenarios. However, in our current study, we still have room for further improvement. For example, our current dataset is not balanced (as shown in Fig. 1) for all performed activities which might lead to the biased results towards majority classes. To handle the said problem to some extent, we used one subject cross-validation process for different models (personal and impersonal) which somehow overcome the class imbalance issue. This study also overcomes the limitations of overlapping while segmenting the raw data which forcefully improve the recognition performance because this overlap cause using some part of data in both training and testing which limiting the use of such techniques in a real-time environment. Moreover, we proposed and suggest some methods which helps to improve the recognition performance for those activities that have low recognition performance.

\begin{figure}[!ht]
	\centering
	\includegraphics[scale=0.16]{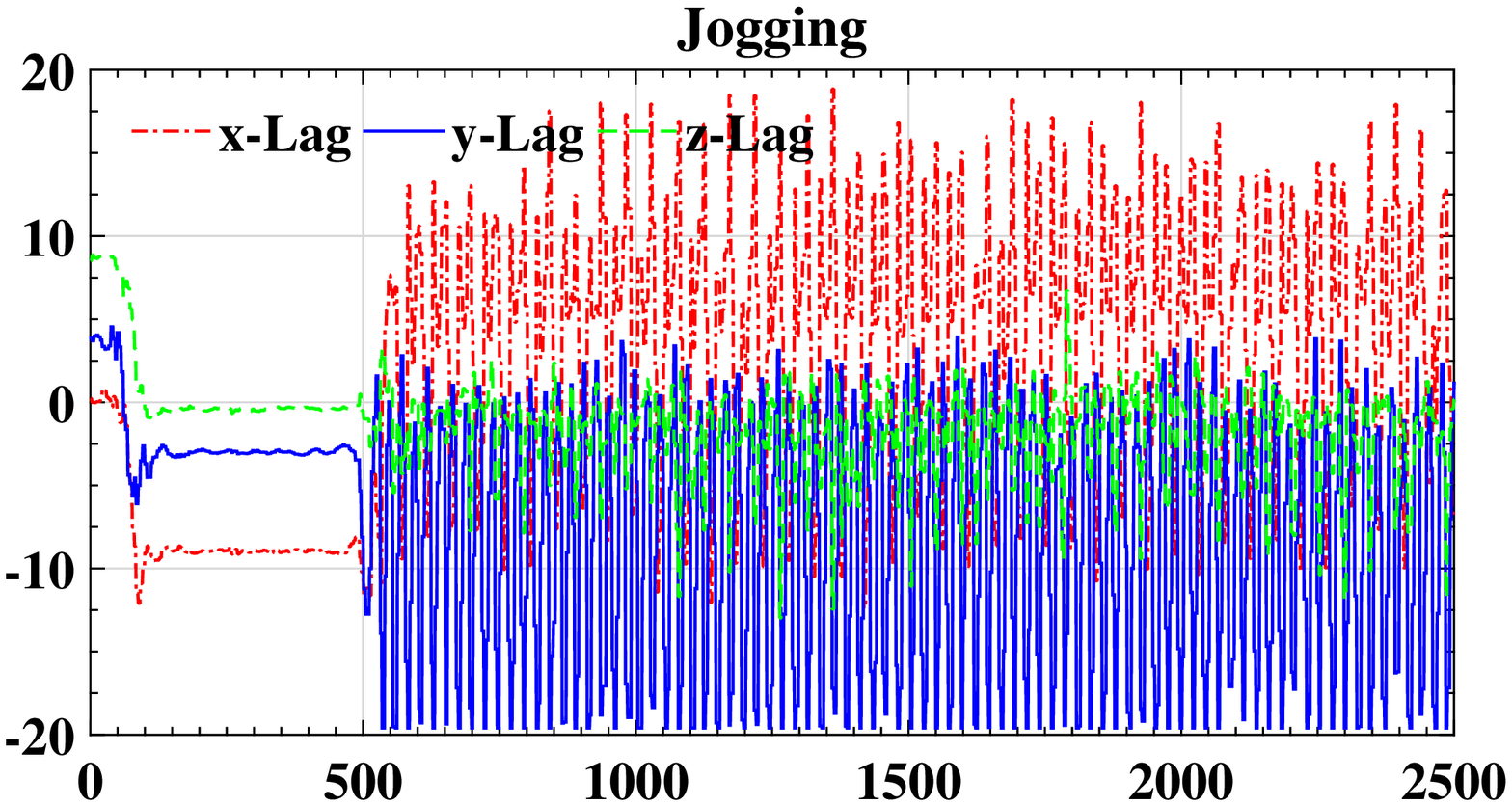}	
	\includegraphics[scale=0.16]{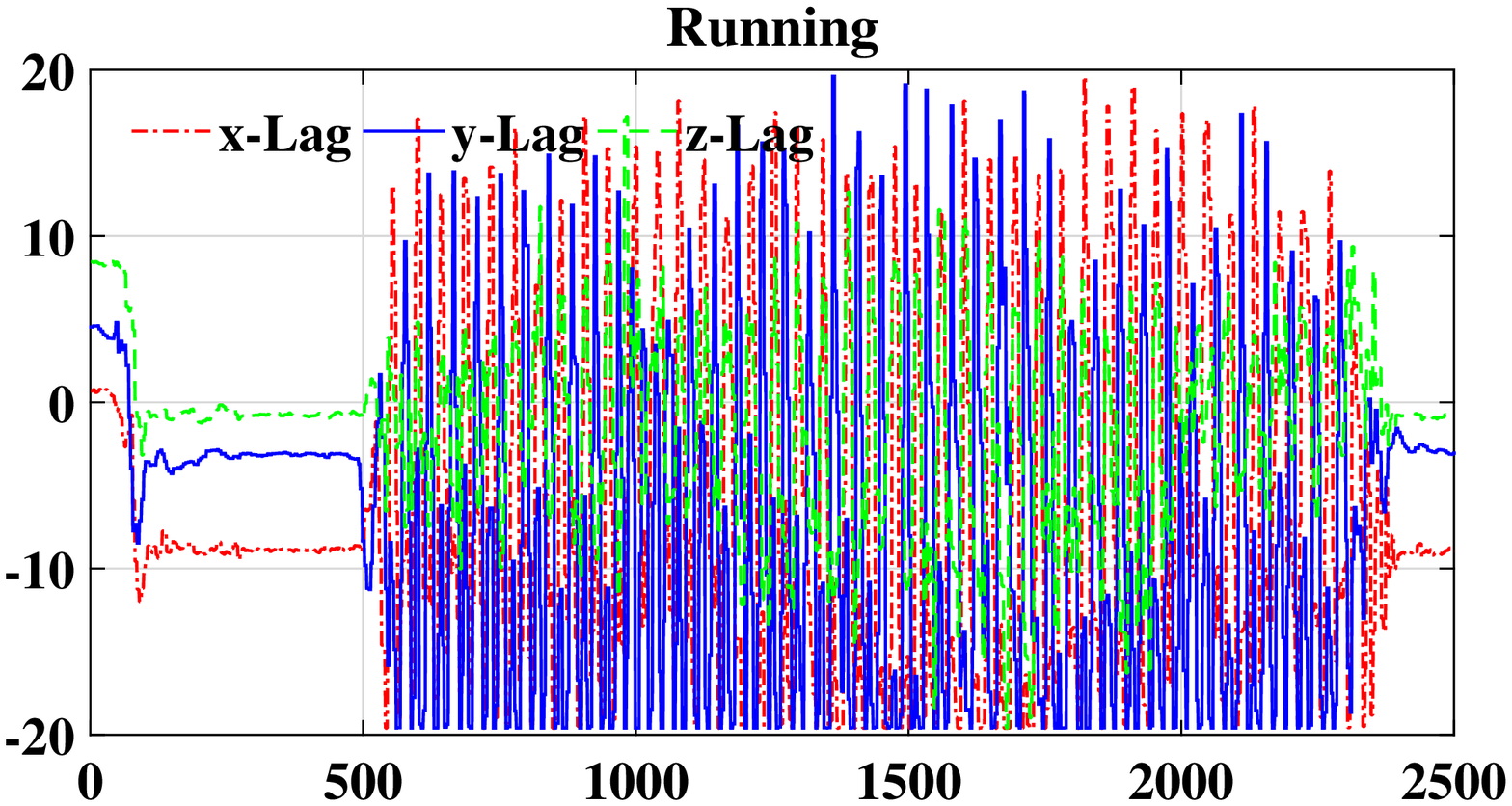}	
	\includegraphics[scale=0.16]{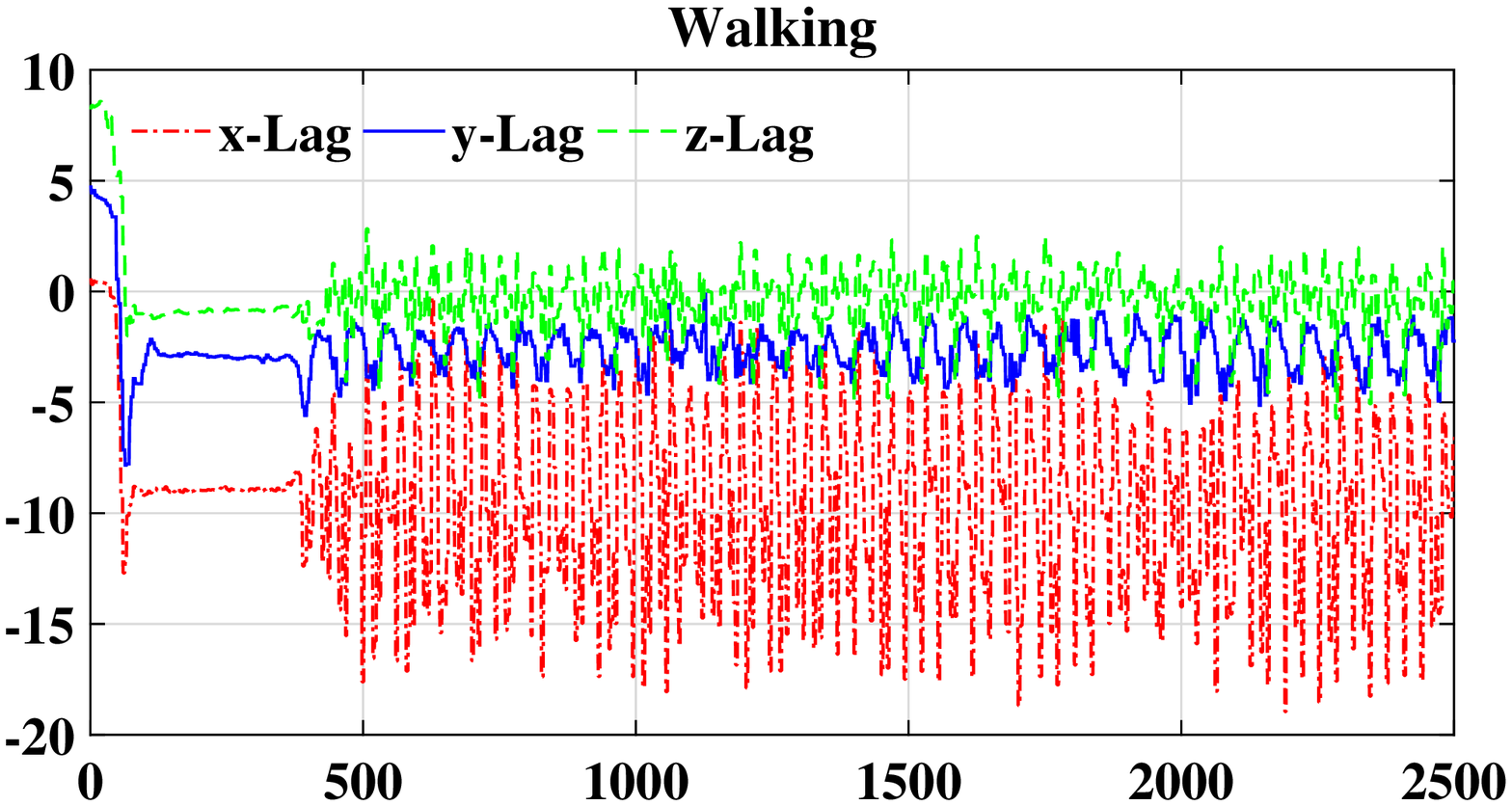} \linebreak
		
	\includegraphics[scale=0.16]{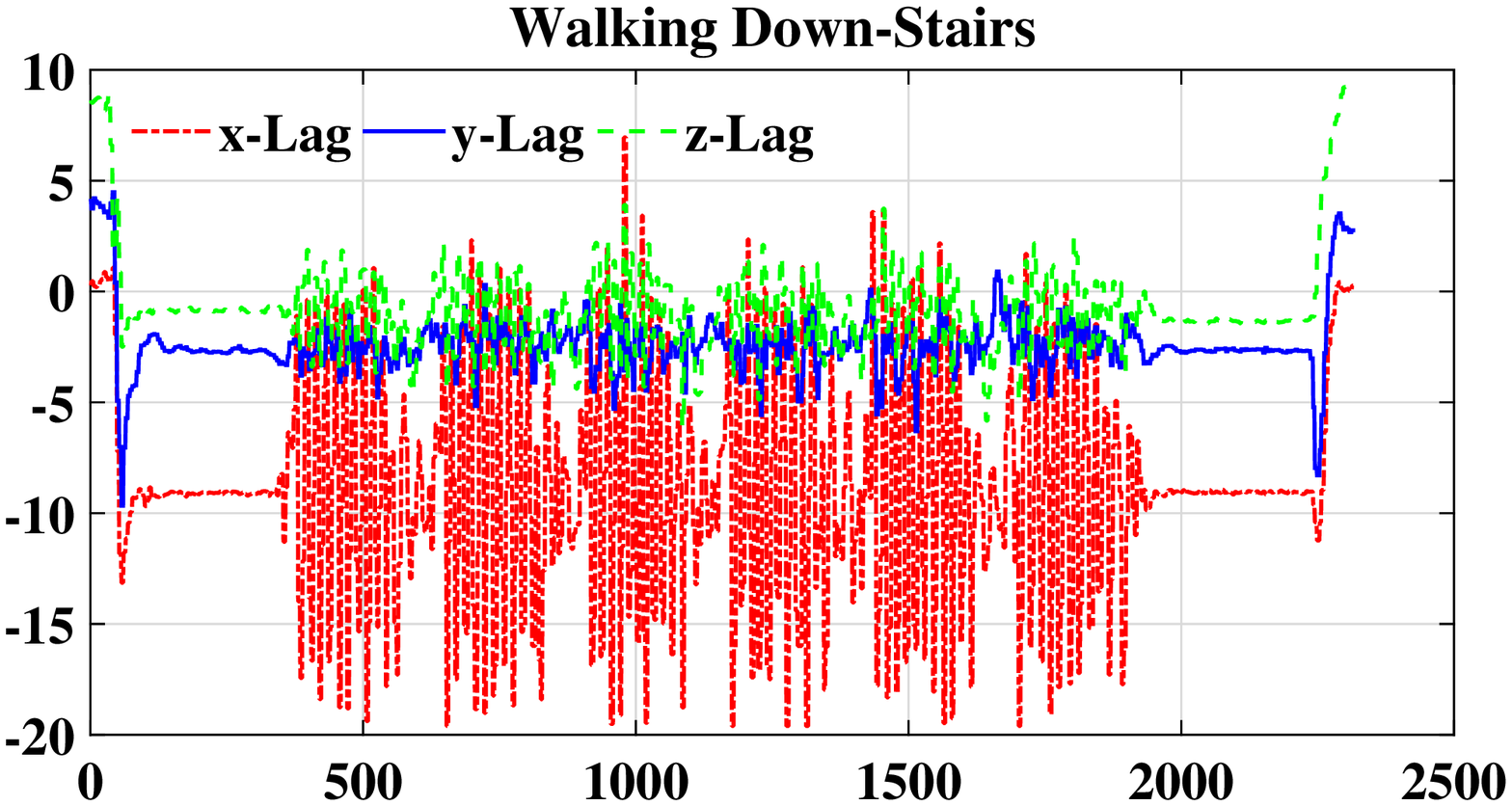}	
	\includegraphics[scale=0.16]{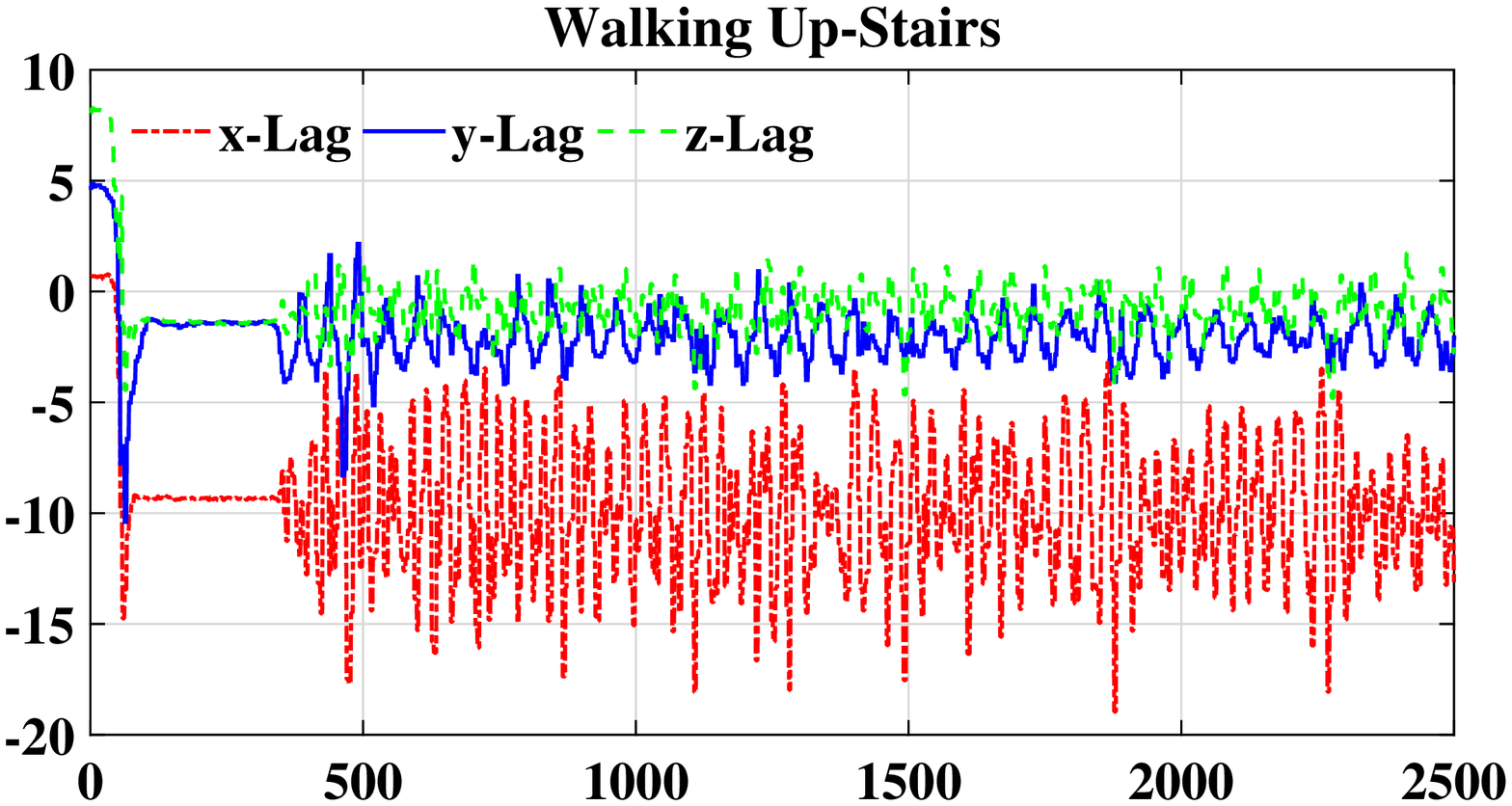}
	\caption{Raw Signal for individual activity.}
\end{figure}

The rest of the paper has the following organization. Section 2 describes the material and methods. Section 3 contains experimental settings. Section 4 presents the experimental results and discussion and finally section 5 concludes the paper with possible future directions.

\section{Materials and Methods}
\label{sec2}

Today's smartwatches are equipped with a variety of motion sensors that are useful for monitoring device movements like tilt, rotate and shake. Some of these sensors are the ambient light sensor, accelerometer, compass, gyroscope and GPS sensors. For experimental study we only collected the raw signals using three different sensors such as accelerometer, magnetometer and gyroscope sensor.

\subsection{Data Collection}
\label{sec3}

We collected the raw signals for five different activities from six users (three female and three male) having a mean age of twenty-five years old. The criterion for selecting the subjects is based on the gender because different genders exhibit different patterns when performing the same activity. These activities include walking, walking upstairs, walking downstairs, running and jogging. All subjects performed these activities twice on each day for more than a month. Therefore we collected raw data from the same user for same activity but performed on different days.

The participants enrolled in our study are approved by the laboratory head because this is a formal prerequisite because our experiments involved human subjects and there is a negligible risk of injury. The involved subjects then asked to answer few nontechnical questions like gender, age, height, weight, left or right handed etc, which we used to characterize our study. Then the subjects were asked to fastens the smartwatch on their wrist and places a Bluetooth paired smartphone in their pocket. Both devices run a simple custom designed application that controls the data collection process and instructs the participant to first add their name and then select the activity from the list of five different activities and the sensor from three different sensors. Once the initial instructions were done, turn the smartphone screen off and place the smartphone into the pants pocket. The smartphone instructs the smartwatch running our paired data collection application to collect the raw signal at the 20Hz rate. Each of these sensors generates 3-dimensional signals and appends a time-stamp to the values. Every after five minutes the smartwatch sends the data to the smartphone and after a successful transmission, the smartphone vibrate to notify the user that the data collection process is successfully completed and they can stop the current activity.

\subsection{Feature Extraction}
\label{sec4}

The embedded sensor generates three-dimensional time series signals which are highly fluctuating and oscillatory in nature. The oscillation and fluctuations make the physical activity recognition more difficult than other applications in nature. These raw signals are shown in Fig. 1 for individual activity performed by the single user. Therefore, it is compulsory to gather the nontrivial signals from raw data through a feature extraction process. The raw signals are divided into several equal sized windows to control the flow rate and pass fewer data to the system to extract the meaningful information. In [24] Khan et al. have presented a detailed analysis of using different kinds of features with a different number of samples per windows for smartphone-based physical activity recognition.

From these raw signals, the extracted features are divided into two different feature banks. First feature bank include the features obtained by taking; average, median, variance, standard deviation, interquartile, autocorrelation, partial autocorrelation, coefficients of autoregressive model, coefficients of moving average model, coefficients of autoregressive moving average model and wavelet coefficients. Second feature bank includes the features obtained by taking; average acceleration, average absolute difference, standard deviation, average resultant acceleration and the average difference between peaks. We also calculate the binned distribution in which we determine what fraction of reading fall in a 10 equal-sized bins. This function generates 10 features. Every feature except average resultant acceleration is extracted for each axis.

These features are extracted from each axis of the three-dimensional acceleration signal and in total for each window of acceleration data 43 (first-feature-bank) and 70 (second-feature-bank) features are extracted respectively. Prior to the feature extraction process moving average filter of order three is used for noise reduction.

\subsection{Classifiers}
\label{sec5}

Khan et al. and Saputri et al. showed that support vector machines (SVM) and artificial neural networks (ANN) are superior to other traditional classification methods for user-independent physical activity recognition system in [24-26]. K-nearest neighbor (KNN) is also one of the most popular classification method used for smartphone-based physical activity recognition and user identification [27]. ELM [28], Naive Bayes [29], Adaptive learning/Ensemble learning (BAG) [30] and Decision tree [31] classifiers are also famous for smartphone-based physical activity recognition system. All these classification methods have shown several advantages in several fields most probably in their respective domains. Once again, given that ours is an exploratory study and the fact that each of these classifiers has shown a good performance in their respective studies. Instead of choosing one we decided to explore the use of all of these classifiers except neural networks in this work.

\section{Experimental Settings}
\label{sec6}

The following experimental studies are performed using all the data to create both personal and impersonal models. In our first study, using a fixed size of 75 samples per window the performance of five classifiers are compared to the following three cases on both feature banks and for both personal and impersonal models: 1): Randomly permuted features with normalization. 2): Without random permutation and normalization. 3): Randomly permuted features without normalization.

In the second study, the best setting for each classifier is chosen and tested multiple times while changing the number of samples in a window \textit{i.e. \{25, 50, 75, 100, 125, 150, 175, 200, 225, 250, 275, 300\}} samples. The goal is to measure the effect of changing the window size on the performance of each classifier for both personal and impersonal model respectively. In our both studies, the training and testing data is randomly divided and classification results are obtained using 10-fold cross-validation process. The values used for different parameters of classifiers are as follows where all these parameters are carefully tuned and optimized. SVM is trained with quadratic kernel function is used, KNN with Euclidean distance is used and K is set to 10, Ensemble model (BAG) with tree model and the number of weak classifiers are set to 50 and the numbers of decision tree are set to 85.

\section{Experimental Results and Discussion}
\label{sec7}

This section contains the obtained results for smartwatch-based physical activity recognition system. To validate our system we conduct different experiments. Our first experiment is for both personal and impersonal model-based physical activity recognition analyses. In this experiment, we conduct a detailed comparison of two different feature banks and five classifiers for both personal and impersonal models. Our second experiment explains the physical activity recognition process behavior with a different number of samples per window within the best settings obtained in our first experiment. Finally, the third experiment presents the single subject cross-validation recognition for each activity for both personal and impersonal models individually. All these experiments are based on 10-fold cross-validation process and all these experiments are carried out using Matlab R2014b installed on core i5 and 8GB of RAM machine.

In the introduction section; we stated some important research questions which our explanatory study seeks to provide the basis of answers. The answers to those questions based on the obtained results are as follows: Based on our findings, yes the features used in existing smartphone-based physical activity recognition studies can also be used for the smartwatch-based physical activity recognition. If we consider normalized and random permutation feature space, classifiers perform much better opposed to the case of without normalization and random permutation in several cases. Table I-IV present the 98\% confidence intervals about overall physical activity recognition rate. Using pairwise T-tests between groups of normalized/randomly permuted and unnormalized/non-permuted data at the 98\% confidence level.

\begin{table*}[!ht]
	\caption{Overall Classification Accuracy and Confidence Interval for \textbf{Impersonal Model} with \textbf{43 Features} extracted using \textbf{75 samples per window}. Where \textit{NR = Normalized, UNR = Unnormalized, RP = Randomly~Permuted, and NRP = Non~Randomly~Permuted}.}
	\resizebox{\textwidth}{!}{\begin{tabular}{l c c c| c c c| c c c| c c c| c c c l}
			\hline \hline
			\multicolumn{1}{c}{\bf Classifiers} & \multicolumn{3}{c}{\bf Walking} & \multicolumn{3}{c}{\bf Walking Up-Stairs} & \multicolumn{3}{c}{\bf Walking Down-Stairs} & \multicolumn{3}{c}{\bf Running}& \multicolumn{3}{c}{\bf Jogging}\\ \hline
			& \multicolumn{1}{c}{\bf NR-RP} & \multicolumn{1}{c}{\bf NR-NRP}& \multicolumn{1}{c}{\bf UNR-RP} & \multicolumn{1}{c}{\bf NR-RP} & \multicolumn{1}{c}{\bf NR-NRP}& \multicolumn{1}{c}{\bf UNR-RP}& \multicolumn{1}{c}{\bf NR-RP} & \multicolumn{1}{c}{\bf NR-NRP}& \multicolumn{1}{c}{\bf UNR-RP}& \multicolumn{1}{c}{\bf NR-RP} & \multicolumn{1}{c}{\bf NR-NRP}& \multicolumn{1}{c}{\bf UNR-RP}& \multicolumn{1}{c}{\bf NR-RP} & \multicolumn{1}{c}{\bf NR-NRP}& \multicolumn{1}{c}{\bf UNR-RP}\\ \hline 
			\bf Decision \\ \bf Tree    & \bf 0.90\(\pm\)0.08 &0.89\(\pm\)0.09	&0.41\(\pm\)0.26	 & \bf 0.90\(\pm\)0.07	& \bf 0.90\(\pm\)0.09 &0.46\(\pm\)0.20	&0.87\(\pm\)0.12	&0.87\(\pm\)0.12	&0.60\(\pm\)0.13 &0.82\(\pm\)0.18	&0.82\(\pm\)0.18	 &0.61\(\pm\)0.15	& \bf 0.93\(\pm\)0.06 	& \bf 0.94\(\pm\)0.06	&0.85\(\pm\)0.10  \\ \hline 
			\bf Naive  \\ \bf Bayes    & \bf 0.86\(\pm\)0.06 & \bf 0.86\(\pm\)0.06&	0.63\(\pm\)0.18&	0.78\(\pm\)0.12&	0.78\(\pm\)0.12&	0.36\(\pm\)0.15&	0.76\(\pm\)0.16&	0.76\(\pm\)0.16& 0.42\(\pm\)0.11&	0.55\(\pm\)0.18&	0.55\(\pm\)0.18&	0.59\(\pm\)0.17&	\bf 0.86\(\pm\)0.02&	\bf 0.86\(\pm\)0.02&	\bf 0.86\(\pm\)0.04  \\ \hline \\
			\bf KNN   & \bf 0.86\(\pm\)0.08&	\bf 0.86\(\pm\)0.08&	0.71\(\pm\)0.08&	0.62\(\pm\)0.07&	0.62\(\pm\)0.07	&0.18\(\pm\)0.07&	0.70\(\pm\)0.13&	0.70\(\pm\)0.13&	0.38\(\pm\)0.12&	0.61\(\pm\)0.17&	0.61\(\pm\)0.17&	0.56\(\pm\)0.13&	 \bf 0.91\(\pm\)0.06&	\bf 0.91\(\pm\)0.06&	\bf 0.88\(\pm\)0.04   \\ \hline \\
			\bf SVM   &\bf 0.94\(\pm\)0.07&	\bf 0.94\(\pm\)0.06&	0.65\(\pm\)0.32&	\bf 0.91\(\pm\)0.04&	\bf 0.91\(\pm\)0.04&	0.25\(\pm\)0.27&	0.81\(\pm\)0.10&	0.81\(\pm\)0.10&	0.43\(\pm\)0.20&	0.76\(\pm\)0.13&	0.76\(\pm\)0.13&	0.55\(\pm\)0.16&	\bf 0.92\(\pm\)0.03&	\bf 0.92\(\pm\)0.03&	\bf 0.88\(\pm\)0.05  \\ \hline \\
			\bf BAG   &\bf 0.91\(\pm\)0.08&	\bf 0.90\(\pm\)0.08&	0.43\(\pm\)0.24& \bf 0.89\(\pm\)0.09&	\bf 0.89\(\pm\)0.09 &	0.42\(\pm\)0.19 &	 \bf 0.87\(\pm\)0.12&	\bf 0.87\(\pm\)0.12 &	0.60\(\pm\)0.13&	0.81\(\pm\)0.18 &  0.81\(\pm\)0.18 &	 0.60\(\pm\)0.16 &	 \bf 0.94\(\pm\)0.05 &	\bf 0.93\(\pm\)0.06 &	\bf 0.85\(\pm\)0.09 \\ \hline 
	\end{tabular}}
\end{table*}

\begin{table*}[!ht]
	\caption{Overall Classification Accuracy and Confidence Interval for \textbf{Personal Model} with \textbf{43 Features} extracted using \textbf{75 samples per window}}
	\resizebox{\textwidth}{!}{\begin{tabular}{l c c c| c c c| c c c| c c c| c c c l}
			\hline \hline
			\multicolumn{1}{c}{\bf Classifiers} & \multicolumn{3}{c}{\bf Walking} & \multicolumn{3}{c}{\bf Walking Up-Stairs} & \multicolumn{3}{c}{\bf Walking Down-Stairs} & \multicolumn{3}{c}{\bf Running}& \multicolumn{3}{c}{\bf Jogging}\\ \hline
			& \multicolumn{1}{c}{\bf NR-RP} & \multicolumn{1}{c}{\bf NR-NRP}& \multicolumn{1}{c}{\bf UNR-RP} & \multicolumn{1}{c}{\bf NR-RP} & \multicolumn{1}{c}{\bf NR-NRP}& \multicolumn{1}{c}{\bf UNR-RP}& \multicolumn{1}{c}{\bf NR-RP} & \multicolumn{1}{c}{\bf NR-NRP}& \multicolumn{1}{c}{\bf UNR-RP}& \multicolumn{1}{c}{\bf NR-RP} & \multicolumn{1}{c}{\bf NR-NRP}& \multicolumn{1}{c}{\bf UNR-RP}& \multicolumn{1}{c}{\bf NR-RP} & \multicolumn{1}{c}{\bf NR-NRP}& \multicolumn{1}{c}{\bf UNR-RP}\\ \hline
			\bf Decision \\ \bf Tree  & \bf 0.98\(\pm\)0.01&	0.88\(\pm\)0.09&	\bf 0.94\(\pm\)0.01&	\bf 0.93\(\pm\)0.02&	0.78\(\pm\)0.06&	0.69\(\pm\)0.10&	\bf 0.98\(\pm\)0.05&	0.75\(\pm\)0.03&	0.75\(\pm\)0.07&	\bf 0.95\(\pm\)0.01&	0.95\(\pm\)0.01&	0.78\(\pm\)0.09&	\bf 0.98\(\pm\)0.01&	\bf 0.95\(\pm\)0.06&	\bf 0.94\(\pm\)0.01  \\ \hline \\
			\bf Naive    & \bf 0.91\(\pm\)0.02&	0.72\(\pm\)0.02&	\bf 0.87\(\pm\)0.08&	0.77\(\pm\)0.11&	0.61\(\pm\)0.09&	0.55\(\pm\)0.07&	0.81\(\pm\)0.07&	0.60\(\pm\)0.08&	0.64\(\pm\)0.12&	0.61\(\pm\)0.19&	0.67\(\pm\)0.13&	0.65\(\pm\)0.07&	\bf 0.87\(\pm\)0.08&	0.73\(\pm\)0.07&	\bf 0.87\(\pm\)0.06  \\ \hline \\
			\bf KNN   & \bf 0.92\(\pm\)0.07&	0.79\(\pm\)0.07&	0.87\(\pm\)0.07&	0.64\(\pm\)0.09&	0.45\(\pm\)0.11&	0.22\(\pm\)0.05&	0.62\(\pm\)0.12&	0.53\(\pm\)0.10&	0.44\(\pm\)0.10&	0.77\(\pm\)0.07&	0.72\(\pm\)0.12&	0.71\(\pm\)0.08&	\bf 0.91\(\pm\)0.08&	\bf 0.88\(\pm\)0.09&	\bf 0.92\(\pm\)0.03   \\ \hline \\
			\bf SVM   & \bf 0.93\(\pm\)0.07&	0.76\(\pm\)0.01&	\bf 0.93\(\pm\)0.01&	0.79\(\pm\)0.09&	0.74\(\pm\)0.07&	0.56\(\pm\)0.16&	\bf 0.90\(\pm\)0.03& 0.68\(\pm\)0.04&	0.6\(\pm\)0.10&	\bf 0.87\(\pm\)0.06&	\bf 0.83\(\pm\)0.09&	0.74\(\pm\)0.07&	\bf 0.97\(\pm\)0.01&	\bf 0.94\(\pm\)0.06&	\bf 0.94\(\pm\)0.01\\ \hline \\
			\bf BAG   & \bf 0.98\(\pm\)0.01&	\bf 0.88\(\pm\)0.09&	\bf 0.87\(\pm\)0.09&	0.85\(\pm\)0.09&	0.80\(\pm\)0.08&	0.66\(\pm\)0.06&	\bf 0.89\(\pm\)0.07&	0.75\(\pm\)0.04&	0.72\(\pm\)0.09&	\bf 0.95\(\pm\)0.02&	\bf 0.95\(\pm\)0.02&	0.72\(\pm\)0.10&	\bf 0.99\(\pm\)0.01&	\bf 0.92\(\pm\)0.08&	\bf 0.94\(\pm\)0.01 \\ \hline \end{tabular}}
\end{table*}

\begin{table*}[!ht]
	\caption{Overall Classification Accuracy and Confidence Interval for \textbf{Personal Model} with \textbf{70 Features} extracted using \textbf{75 samples per window}}
	\resizebox{\textwidth}{!}{\begin{tabular}{l c c c| c c c| c c c| c c c| c c c l}
			\hline \hline
			\multicolumn{1}{c}{\bf Classifiers} & \multicolumn{3}{c}{\bf Walking} & \multicolumn{3}{c}{\bf Walking Up-Stairs} & \multicolumn{3}{c}{\bf Walking Down-Stairs} & \multicolumn{3}{c}{\bf Running}& \multicolumn{3}{c}{\bf Jogging}\\ \hline
			& \multicolumn{1}{c}{\bf NR-RP} & \multicolumn{1}{c}{\bf NR-NRP}& \multicolumn{1}{c}{\bf UNR-RP} & \multicolumn{1}{c}{\bf NR-RP} & \multicolumn{1}{c}{\bf NR-NRP}& \multicolumn{1}{c}{\bf UNR-RP}& \multicolumn{1}{c}{\bf NR-RP} & \multicolumn{1}{c}{\bf NR-NRP}& \multicolumn{1}{c}{\bf UNR-RP}& \multicolumn{1}{c}{\bf NR-RP} & \multicolumn{1}{c}{\bf NR-NRP}& \multicolumn{1}{c}{\bf UNR-RP}& \multicolumn{1}{c}{\bf NR-RP} & \multicolumn{1}{c}{\bf NR-NRP}& \multicolumn{1}{c}{\bf UNR-RP}\\ \hline 
			\bf Decision \\ \bf Tree  &\bf 0.98\(\pm\)0.01& \bf 0.87\(\pm\)0.09& \bf 0.90\(\pm\)0.08& 0.84\(\pm\)0.09& 0.67\(\pm\)0.04& 0.62\(\pm\)0.13& 0.79\(\pm\)0.11& 0.74\(\pm\)0.12& 0.74\(\pm\)0.04& \bf 0.86\(\pm\)0.07& \bf 0.87\(\pm\)0.06& 0.72\(\pm\)0.07& \bf 0.97\(\pm\)0.01& \bf 0.94\(\pm\)0.07& \bf 0.93\(\pm\)0.01    \\ \hline \\
			\bf Naive    & \bf 0.88\(\pm\)0.02& 0.75\(\pm\)0.08& 0.77\(\pm\)0.09& 0.65\(\pm\)0.16& 0.51\(\pm\)0.10& 0.53\(\pm\)0.12& 0.62\(\pm\)0.15& 0.47\(\pm\)0.09& 0.60\(\pm\)0.08& 0.58\(\pm\)0.12& 0.55\(\pm\)0.06&
			0.65\(\pm\)0.14& \bf 0.88\(\pm\)0.02& 0.68\(\pm\)0.02& 0.83\(\pm\)0.07 \\ \hline \\
			\bf KNN  &\bf 0.96\(\pm\) 0.01& \bf 0.93\(\pm\) 0.07&	0.79\(\pm\)	0.09&	0.60\(\pm\)	0.10& 0.46\(\pm\) 0.08& 0.19\(\pm\) 0.11& 0.49\(\pm\) 0.16& 0.46\(\pm\) 0.10& 0.17\(\pm\) 0.06& 0.58±\(\pm\) 0.14& 0.59\(\pm\) 0.16& 0.26\(\pm\) 0.14& \bf 0.95\(\pm\) 0.02& \bf 0.95\(\pm\) 0.02& \bf 0.90\(\pm\) 0.02 \\ \hline \\
			\bf SVM &0.98\(\pm\)0.01&0.91\(\pm\)0.08&0.87\(\pm\)0.07&0.89\(\pm\)0.08&0.77\(\pm\)0.08&0.52\(\pm\)0.07&0.86\(\pm\)0.08&0.76\(\pm\)0.08&0.54\(\pm\)0.08&0.91\(\pm\)0.02&0.85\(\pm\)0.09&
			0.66\(\pm\)0.09&0.97\(\pm\)0.01&0.90\(\pm\)0.10&0.92\(\pm\)0.01\\ \hline \\ 
			\bf BAG   &\bf 0.97\(\pm\)0.01& \bf 0.87\(\pm\)0.09& \bf 0.90\(\pm\)0.07& \bf0.86\(\pm\)0.07& 0.69\(\pm\)0.04& 0.68\(\pm\)0.10& 0.84\(\pm\)0.05& 0.65\(\pm\)0.09& 0.65\(\pm\)0.06& \bf 0.90\(\pm\)0.03& \bf 0.86\(\pm\)0.05& 0.72\(\pm\)0.08& \bf 0.97\(\pm\)0.01& \bf 0.87\(\pm\)0.10& \bf 0.91\(\pm\)0.07 \\ \hline 
	\end{tabular}}
\end{table*}

\begin{table*}[!ht]
	\caption{Overall Classification Accuracy and Confidence Interval for \textbf{Impersonal Model} with \textbf{70 Features} extracted using \textbf{75 samples per window}.}
	\resizebox{\textwidth}{!}{\begin{tabular}{l c c c| c c c| c c c| c c c| c c c l}
			\hline \hline
			\multicolumn{1}{c}{\bf Classifiers} & \multicolumn{3}{c}{\bf Walking} & \multicolumn{3}{c}{\bf Walking Up-Stairs} & \multicolumn{3}{c}{\bf Walking Down-Stairs} & \multicolumn{3}{c}{\bf Running}& \multicolumn{3}{c}{\bf Jogging}\\ \hline
			& \multicolumn{1}{c}{\bf NR-RP} & \multicolumn{1}{c}{\bf NR-NRP}& \multicolumn{1}{c}{\bf UNR-RP} & \multicolumn{1}{c}{\bf NR-RP} & \multicolumn{1}{c}{\bf NR-NRP}& \multicolumn{1}{c}{\bf UNR-RP}& \multicolumn{1}{c}{\bf NR-RP} & \multicolumn{1}{c}{\bf NR-NRP}& \multicolumn{1}{c}{\bf UNR-RP}& \multicolumn{1}{c}{\bf NR-RP} & \multicolumn{1}{c}{\bf NR-NRP}& \multicolumn{1}{c}{\bf UNR-RP}& \multicolumn{1}{c}{\bf NR-RP} & \multicolumn{1}{c}{\bf NR-NRP}& \multicolumn{1}{c}{\bf UNR-RP}\\ \hline 
			\bf Decision \\ \bf Tree  &\bf 0.91\(\pm\)0.07&\bf 0.91\(\pm\)0.07& 0.50\(\pm\)0.21& 0.82\(\pm\)0.09& 0.81\(\pm\)0.09& 0.45\(\pm\)0.19& 0.64\(\pm\)0.12& 0.64\(\pm\)0.12& 0.57\(\pm\)0.10& 0.62\(\pm\)0.21& 0.63\(\pm\)0.21& 0.54\(\pm\)0.18& \bf 0.86\(\pm\)0.07& \bf 0.86\(\pm\)0.07& \bf 0.85\(\pm\)0.094   \\ \hline \\
			\bf Naive    &0.76\(\pm\)0.09& 0.73\(\pm\)0.08& 0.39\(\pm\)0.11& 0.54\(\pm\)0.08& 0.14\(\pm\)0.03& 0.50\(\pm\)0.07& 0.53\(\pm\)0.11& 0.13\(\pm\)0.03& 0.46\(\pm\)0.07& 0.42\(\pm\)0.10& 0.16\(\pm\)0.07& 0.54\(\pm\)0.17& 0.81\(\pm\)0.05& 0.80\(\pm\)0.11& 0.80\(\pm\)0.08  \\ \hline \\
			\bf KNN &\bf 0.89\(\pm\)0.05& \bf 0.89\(\pm\)0.05& 0.73\(\pm\)0.08& 0.54\(\pm\)0.08& 0.54\(\pm\)0.08& 0.14\(\pm\)0.03& 0.42\(\pm\)0.13& 0.42\(\pm\)0.13& 0.13\(\pm\)0.03& 0.31\(\pm\)0.11& 0.31\(\pm\)0.11& 0.16\(\pm\)0.07& \bf 0.86\(\pm\)0.06& \bf 0.88\(\pm\)0.09& 0.80\(\pm\)0.11   \\ \hline \\
			\bf SVM   &\bf 0.97\(\pm\)0.02& \bf 0.97\(\pm\)0.02& 0.84\(\pm\)0.12& \bf 0.90\(\pm\)0.04& \bf 0.90\(\pm\)0.04& 0.12\(\pm\)0.07& 0.82\(\pm\)0.06& 0.82\(\pm\)0.06& 0.50\(\pm\)0.1& 0.80\(\pm\)0.09& 0.80\(\pm\)0.09& 0.60\(\pm\)0.06& \bf 0.94\(\pm\)0.03& \bf 0.94\(\pm\)0.03& \bf 0.87\(\pm\)0.06 \\ \hline \\
			\bf BAG   &\bf 0.90\(\pm\)0.07& \bf 0.90\(\pm\)0.08& 0.51\(\pm\)0.20& 0.80\(\pm\)0.09& 0.80\(\pm\)0.10& 0.44\(\pm\)0.18& 0.63\(\pm\)0.12& 0.63\(\pm\)0.12& 0.57\(\pm\)0.09& 0.61\(\pm\)0.21& 0.61\(\pm\)0.21& 0.54\(\pm\)0.18& 0\bf 0.86\(\pm\).07& \bf 0.86\(\pm\)0.07& \bf 0.85\(\pm\)0.09 \\ \hline 
	\end{tabular}}
\end{table*}

Statistically significant results are in boldface, shows decision trees and SVM classifiers are statistically better when normalized and randomly permuted features are used. This increase was as much as almost 50\% (from 40\% to 90\% for the impersonal model and 94\% to 98\% for the personal model) looking at Table I and Table II with 43 features. Other classifiers are also statistically better using normalized features in a number of cases. For 70 number of feature-bank we again used pairwise T-tests between groups of normalized/randomly permuted and unnormalized/non-permuted data at 97\% - 98\% confidence level.

Statistically significant results are in boldface shows in all cases again decision trees and SVM classifiers are statistically better when normalized features are used. This increase was as much as almost 10\% (from 87\% to 97\% for the personal model) looking at Table III-IV. Table V-VIII shows single subject cross-validation process for each activity recognition.

\begin{figure}[!ht]
	\centering{\includegraphics[scale=0.18]{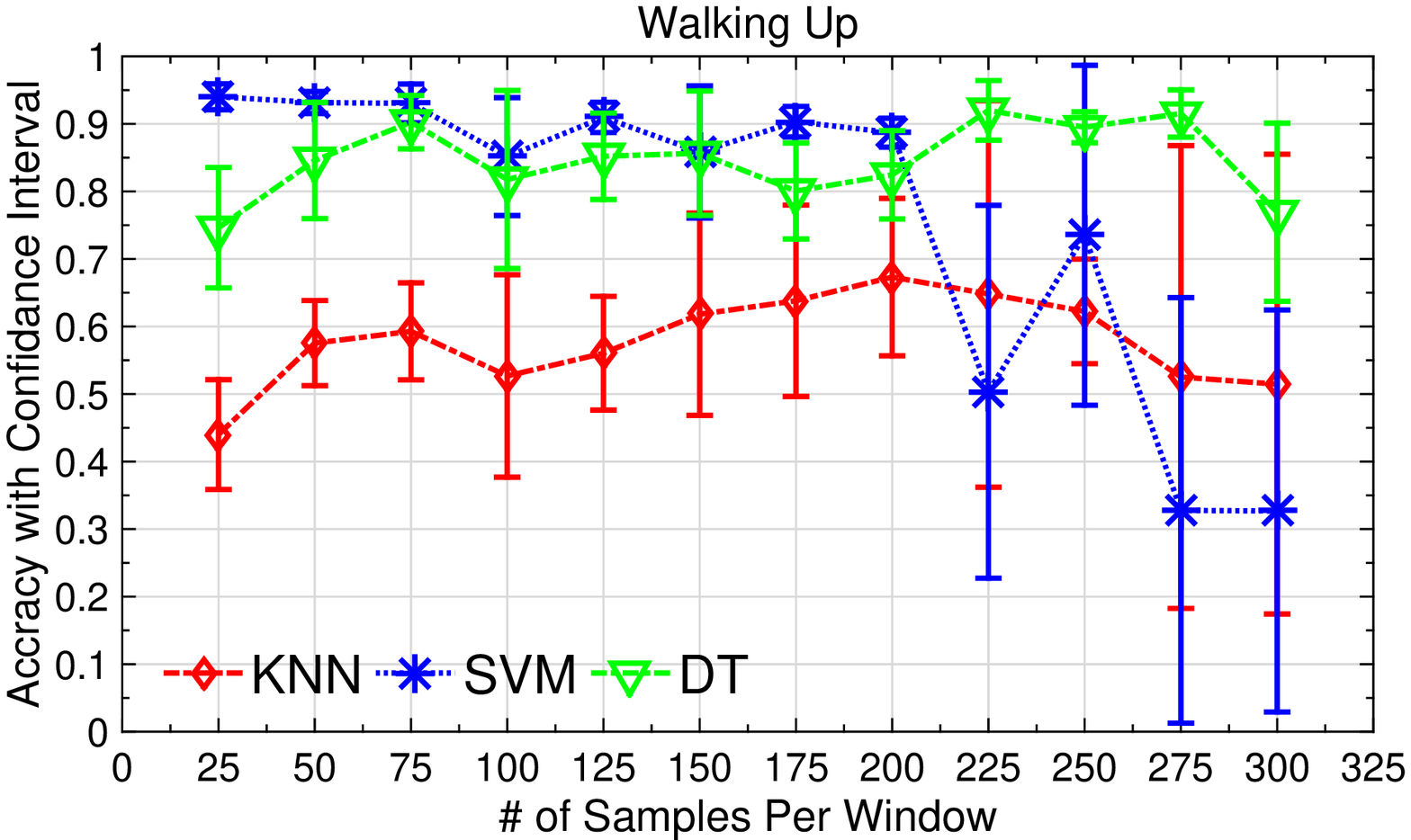}}	
	\centering{\includegraphics[scale=0.18]{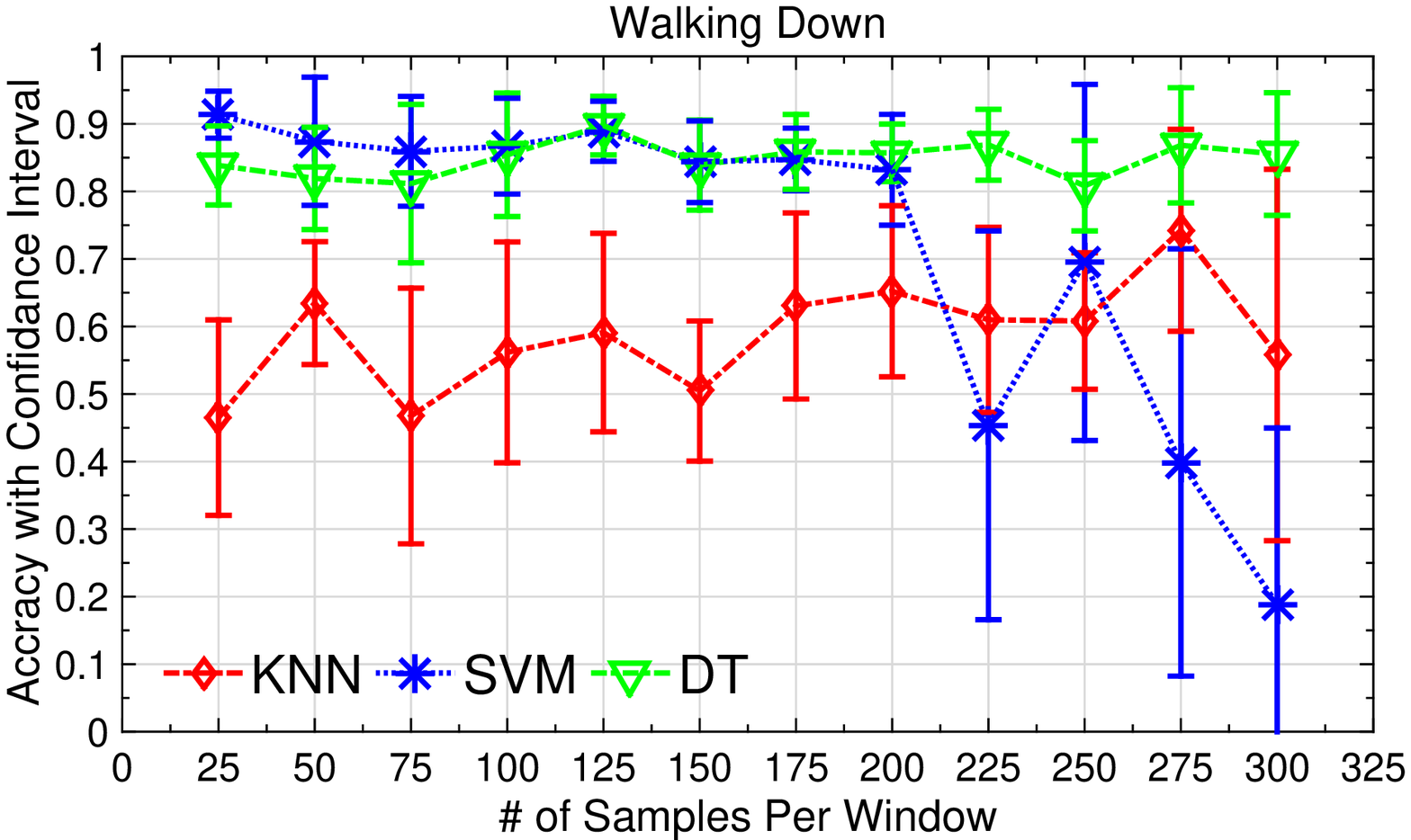}}	
	\centering{\includegraphics[scale=0.18]{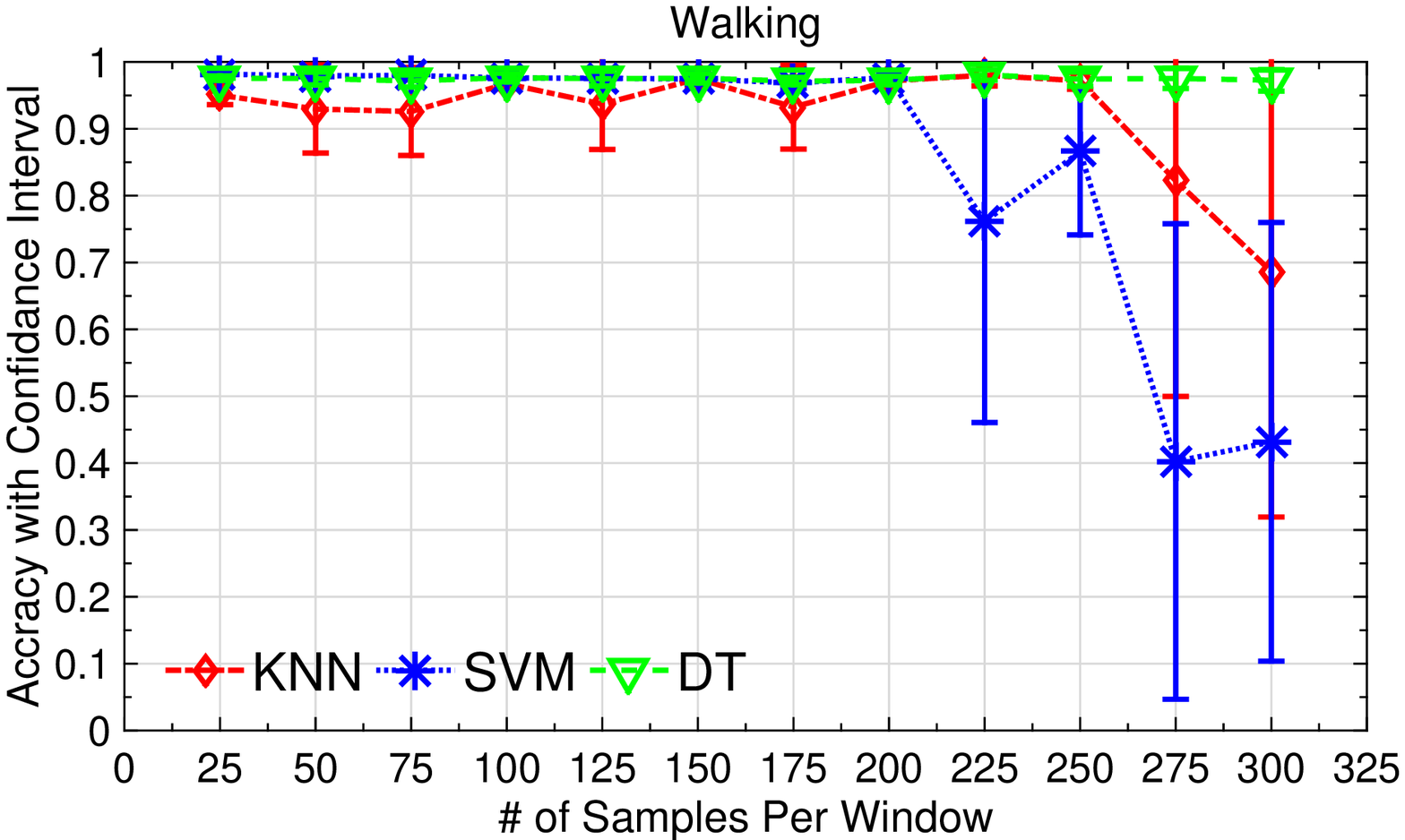}}	\linebreak
	
	\centering{\includegraphics[scale=0.18]{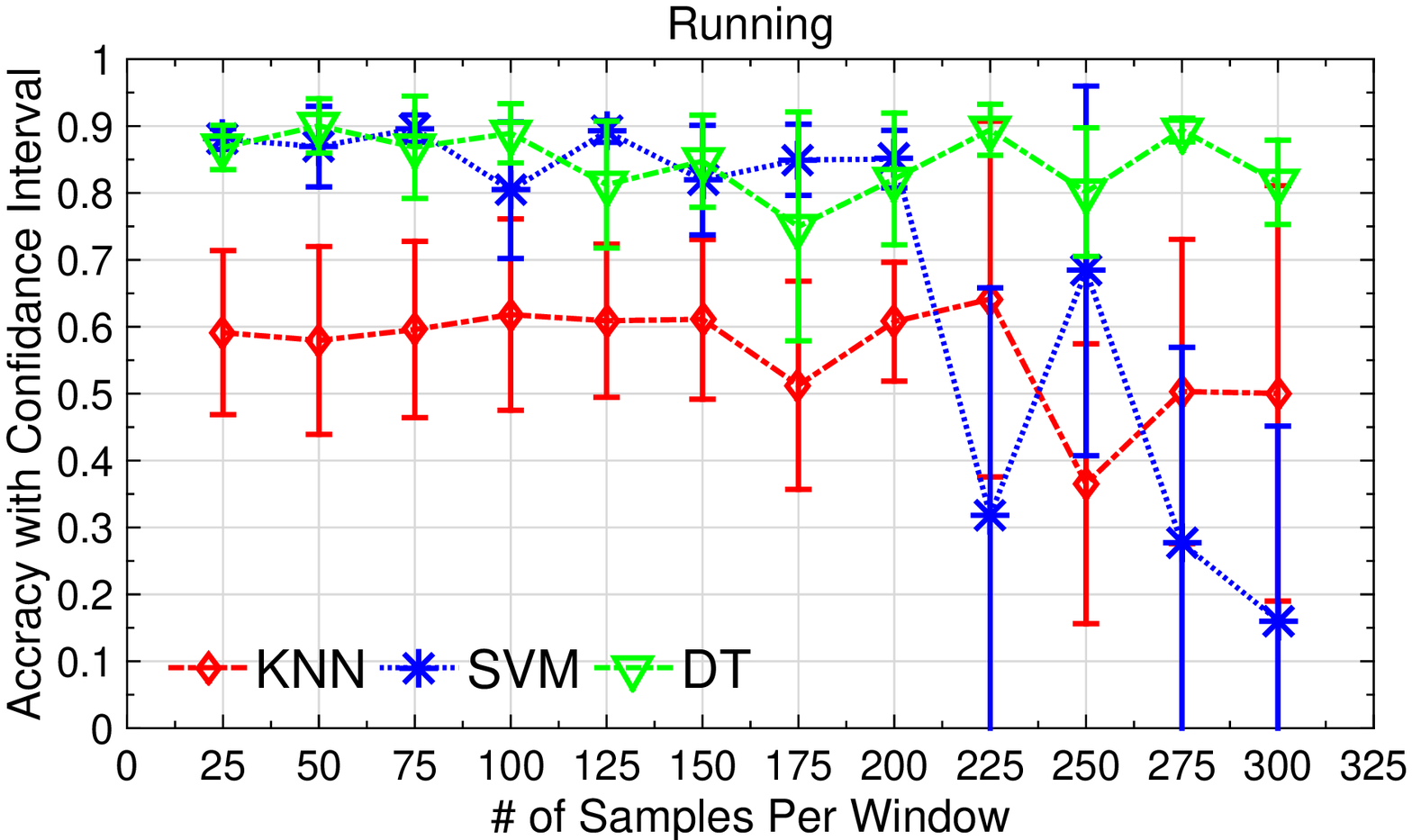}}	
	\centering{\includegraphics[scale=0.18]{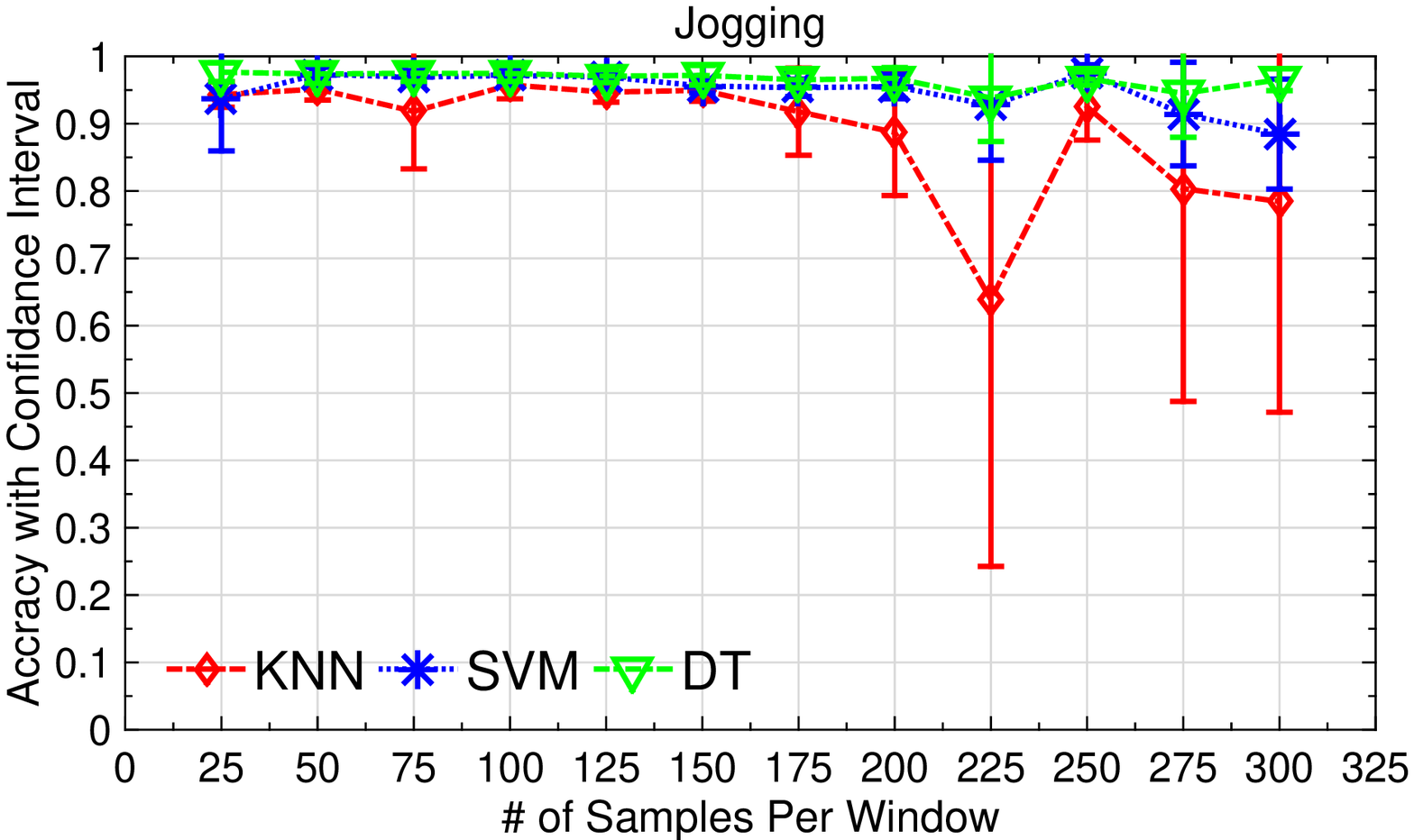}}
	\caption{Overall Recognition Accuracy for different Activities with different number of samples per window.}
\end{figure}

Examining the overall and single subject cross-validation accuracy of all classifiers, each personal and the impersonal model becomes statistically significant in a number of cases. This leads us to prefer the use of normalization and random permutation in future applications of this method and in practical implementations. All of our chosen classifiers provide acceptable performance but decision tree and SVM are found to be better than other classifiers and have a number of other attractive features to their applicability. 

In terms of the applicability of the smartwatch system, decision tree and SVM have smaller confidence intervals implying that they have more reliability in their training than other classification models. In terms of the applicability of the smartwatch system, the personal model has smaller confidence intervals implying that personal model has more reliability overall accuracies for both numbers of feature banks.

\begin{table}[!ht]
	\tiny
	\caption{Single Subject cross-validation-based accuracy analysis for \textbf{Impersonal Model with 43 Features extracted using 75 samples per window}.}
	\resizebox{\textwidth}{!}{\begin{tabular}{l c c c c c c c c c c c c c c c c}
		\hline \hline
		\multicolumn{1}{c}{\bf Users} & \multicolumn{3}{|c|}{\bf Walking} & \multicolumn{3}{c|}{\bf Walking Up-Stairs} & \multicolumn{3}{c|}{\bf Walking Down-Stairs} & \multicolumn{3}{c|}{\bf Running}& \multicolumn{3}{c}{\bf Jogging}\\ \hline
		& \multicolumn{1}{|c}{\bf NR-RP} & \multicolumn{1}{c}{\bf NR-NRP}& \multicolumn{1}{c|}{\bf UNR-RP} & \multicolumn{1}{c}{\bf NR-RP} & \multicolumn{1}{c}{\bf NR-NRP}& \multicolumn{1}{c|}{\bf UNR-RP}& \multicolumn{1}{c}{\bf NR-RP} & \multicolumn{1}{c}{\bf NR-NRP}& \multicolumn{1}{c|}{\bf UNR-RP}& \multicolumn{1}{c}{\bf NR-RP} & \multicolumn{1}{c}{\bf NR-NRP}& \multicolumn{1}{c|}{\bf UNR-RP}& \multicolumn{1}{c}{\bf NR-RP} & \multicolumn{1}{c}{\bf NR-NRP}& \multicolumn{1}{c}{\bf UNR-RP}\\ \hline  \hline
		\multicolumn{16}{c}{\bf Decision Trees Classifier} \\ \hline \hline
		\bf User 1  &0.8706&0.8617&0.0595&0.9714&0.9689&0.4646& 0.7909&0.7942&0.8123& 0.6807&0.6922 &0.5766 &0.9759& 0.9765& 0.9175   \\ \hline
		\bf User 2  &0.7145&0.6961&0.0839&0.9768&0.9794&0.7112&0.9944& 0.9940&0.7330& 0.9724& 0.9688& 0.7371& 0.9844& 0.9871&0.8945   \\ \hline
		\bf User 3  &0.9757&0.9761&0.3730&0.7431&0.7000&0.7931&0.9497&0.9492&0.4230&0.9274& 0.9259& 0.7513&0.8691&0.8864&0.6500   \\ \hline
		\bf User 4  &0.9935 &0.9894 &0.5847&0.9380&0.9390&0.2432&0.9277&0.9312&0.4050&0.4287&0.4201&0.2616&0.9911&0.9941&0.9410   \\ \hline
		\bf User 5  &0.8998&0.8825&0.8987&0.9500&0.9442&0.2238&0.9690&0.9665&0.6171&0.9164&0.9126&0.7327&0.9687&0.9646&0.9400   \\ \hline
		\bf User 6  & 0.9563&0.9518&0.4594&0.8450&0.8462&0.2972&0.6004&0.6025&0.6017&0.9804&0.9812&0.5938&0.8039&0.7993&0.7626 \\ \hline \hline
		\multicolumn{16}{c}{\bf Naive Bayes Classifier} \\ \hline \hline
		\bf User 1  &0.7597& 0.7597& 0.2070& 0.8883&  0.8883& 0.3495& 0.6883& 0.6883& 0.5390&  0.2448&  0.2448& 0.4323& 0.8522& 0.8522& 0.8595   \\ \hline
		\bf User 2  &0.7979&  0.7979 & 0.6905&  0.9270&  0.9270& 0.3863& 0.9742 &0.9742 &0.6137 &0.7647 &0.7647& 0.7235 &0.8855 &0.8855 &0.8909   \\ \hline
		\bf User 3  &0.9179 &0.9179 &0.6589 &0.6373 &0.6373& 0.6961 &0.8907 &0.8907 &0.3060 &0.7411& 0.7411& 0.7614& 0.8287& 0.8287 &0.7832   \\ \hline
		\bf User 4  &0.9369 &0.9369 &0.8342  &0.8545& 0.8545& 0.2723 &0.7525 &0.7525 &0.2624 &0.3171& 0.3171& 0.2317& 0.8614& 0.8614 &0.9242  \\ \hline
		\bf User 5  &0.8667 &0.8667 &0.5657&0.7791 &0.7791 &0.2500& 0.8734 &0.8734 &0.4430 &0.6918 &0.6918 &0.7673& 0.9038&  0.9038& 0.8308   \\ \hline
		\bf User 6  & 0.8847& 0.8847& 0.8176& 0.5663& 0.5663& 0.1847& 0.4017& 0.4017& 0.3473 &0.5134 &0.5134 &0.6205& 0.8256& 0.8256& 0.8479 \\ \hline \hline
		\multicolumn{16}{c}{\bf KNN Classifier} \\ \hline \hline
		\bf User 1  &0.7523 &0.7523& 0.6728 &0.5680& 0.5680& 0.1359 &0.6299 &0.6299 &0.4416 &0.4688 &0.4688 &0.5365& 0.9179 &0.9179 &0.8942   \\ \hline
		\bf User 2  &0.7370& 0.7370&  0.6279&  0.7854& 0.7854& 0.2017& 0.8970& 0.8970 &0.3605  &0.7765 &0.7765 &0.6882& 0.9655 &0.9655& 0.9091   \\ \hline
		\bf User 3  &0.9661 &0.9661 &0.8268 &0.6618 &0.6618 &0.3235 &0.7814 &0.7814 &0.1913 &0.7310 &0.7310 &0.6853& 0.8759& 0.8759& 0.8444  \\ \hline
		\bf User 4  &0.8901 &0.8901  &0.6811 &0.6150 &0.6150 &0.1878& 0.7030 &0.7030 &0.2327 &0.2317 &0.2317  &0.2378& 0.9778& 0.9778& 0.9298  \\ \hline
		\bf User 5  &0.9371 &0.9371 &0.8495 &0.5698 &0.5698 &0.0698 &0.7848 &0.7848 &0.5190 &0.6667 &0.6667& 0.5723 & 0.9615 &0.9615& 0.9269   \\ \hline
		\bf User 6  &0.8985 &0.8985& 0.6127& 0.5422 &0.5422 &0.1606 &0.4184& 0.4184 &0.5565& 0.7813& 0.7813& 0.6339& 0.7658 &0.7658& 0.8051 \\ \hline \hline
		\multicolumn{16}{c}{\bf SVM Classifier} \\ \hline \hline
		\bf User 1  &0.9760 &0.9760 &0.1257 &0.9223 &0.9223 &0.3204& 0.8377 &0.9657 &0.6883 &0.4740 &0.4740 &0.4792& 0.9398& 0.9380& 0.9033  \\ \hline
		\bf User 2  &0.7710 &0.7710 &0.8497 &0.9742 &0.9742& 0.0172 &0.9657 &0.8361 &0.6695& 0.8706& 0.8706& 0.7059& 0.9164& 0.9164 &0.9073  \\ \hline
		\bf User 3  &0.9679 &0.9679 &0.1304 &0.8137 &0.8137 &0.8971 &0.8361 &0.7277 &0.2787& 0.8985& 0.8985& 0.6954& 0.8654& 0.8654& 0.7815  \\ \hline
		\bf User 4  &0.9874 &0.9874& 0.9550& 0.9061& 0.9061& 0.0329& 0.7277& 0.8608& 0.2772& 0.7378& 0.7378& 0.1707& 0.9464& 0.9464& 0.9575  \\ \hline
		\bf User 5  &0.9600 &0.9600 & 0.9448 &0.9302 &0.9302 &0.0814 &0.8608 &0.6109 &0.5823& 0.6792& 0.6792 &0.6604& 0.9788 &0.9788& 0.9192    \\ \hline
		\bf User 6  &0.9776 &0.9776& 0.8864& 0.9197& 0.9197& 0.1365& 0.6109 &0.9657 &0.0669 &0.8839 &0.8839 &0.5848& 0.8991 &0.8974 &0.8308 \\ \hline \hline
		\multicolumn{16}{c}{\bf Ensemble Learning-BAG Classifier} \\ \hline \hline
		\bf User 1  &0.8784 &0.8656  &0.0980& 0.9694& 0.9670& 0.3529 &0.7805& 0.7864& 0.7968& 0.6734& 0.6786 &0.5661& 0.9763 &0.9748& 0.9224  \\ \hline
		\bf User 2  &0.7086& 0.7038 &0.1243 & 0.9712& 0.9721 &0.6236 &0.9944 &0.9953  &0.7416 &0.9641 &0.9647 &0.7429& 0.9827& 0.9811& 0.8982  \\ \hline
		\bf User 3  &0.9757& 0.9759 &0.3900 &0.6730& 0.6985& 0.7902& 0.9475 &0.9497 &0.4240 &0.9279& 0.9228& 0.7472& 0.8692 &0.8764 &0.6458  \\ \hline
		\bf User 4  &0.9894 &0.9874 & 0.5782 &0.9376 &0.9366 &0.2531  &0.9292 &0.9233 &0.4183& 0.3988 &0.4073& 0.2317& 0.9930 &0.9904& 0.9442  \\ \hline
		\bf User 5  &0.8989 &0.8970 &0.9006 &0.9517& 0.9448& 0.2017& 0.9614& 0.9608& 0.6253& 0.8994 &0.9031& 0.7409& 0.9677 &0.9662 &0.9448    \\ \hline
		\bf User 6  &0.9546& 0.9497&  0.4305&  0.8454 &0.8426 &0.2924  &0.6167& 0.6079 &0.6004& 0.9772& 0.9777& 0.5960& 0.8373& 0.8113& 0.7576 \\ \hline \hline
	\end{tabular}}
\end{table}

\begin{table}[!ht]
	\tiny
	\caption{Single Subject cross-validation-based accuracy analysis for \textbf{Personal Model with 43 Features extracted using 75 samples per window}.}
	\resizebox{\textwidth}{!}{\begin{tabular}{l c c c c c c c c c c c c c c c c}
		\hline \hline
		\multicolumn{1}{c}{\bf Users} & \multicolumn{3}{|c|}{\bf Walking} & \multicolumn{3}{c|}{\bf Walking Up-Stairs} & \multicolumn{3}{c|}{\bf Walking Down-Stairs} & \multicolumn{3}{c|}{\bf Running}& \multicolumn{3}{c}{\bf Jogging}\\ \hline
		& \multicolumn{1}{|c}{\bf NR-RP} & \multicolumn{1}{c}{\bf NR-NRP}& \multicolumn{1}{c|}{\bf UNR-RP} & \multicolumn{1}{c}{\bf NR-RP} & \multicolumn{1}{c}{\bf NR-NRP}& \multicolumn{1}{c|}{\bf UNR-RP}& \multicolumn{1}{c}{\bf NR-RP} & \multicolumn{1}{c}{\bf NR-NRP}& \multicolumn{1}{c|}{\bf UNR-RP}& \multicolumn{1}{c}{\bf NR-RP} & \multicolumn{1}{c}{\bf NR-NRP}& \multicolumn{1}{c|}{\bf UNR-RP}& \multicolumn{1}{c}{\bf NR-RP} & \multicolumn{1}{c}{\bf NR-NRP}& \multicolumn{1}{c}{\bf UNR-RP}\\ \hline  \hline
		\multicolumn{16}{c}{\bf Decision Trees Classifier} \\ \hline \hline
		\bf User 1  &0.9815 &0.7889 &0.9500 &0.9073 &0.7366 &0.8000 &0.9067 &0.7067 &0.5867 &0.9632 &0.9368 &0.7895& 0.9908& 0.9945& 0.9633   \\ \hline
		\bf User 2  &0.9820 &0.9802 &0.9604  &0.8957 &0.7348 &0.8217 &0.7870& 0.7826& 0.8043& 0.9471& 0.9647 &0.8765 & 0.9836& 0.9873& 0.9382  \\ \hline
		\bf User 3  &0.9786 &0.7768 &0.9214& 0.9450& 0.7500& 0.6750 &0.9278 &0.7167 &0.7611& 0.9333& 0.9128& 0.5487& 0.9772 &0.9579 &0.9193   \\ \hline
		\bf User 4  &0.9874 &0.9946 &0.9568 &0.9714& 0.7619& 0.5000& 0.9600& 0.7400& 0.8100& 0.9500& 0.9625 &0.7875& 0.9833 &0.9815 &0.9500   \\ \hline
		\bf User 5  &0.9867 &0.9905& 0.9314&  0.9235& 0.7412& 0.7176& 0.9226& 0.7290& 0.8129& 0.9548& 0.9484& 0.8581& 0.9865 &0.7885 &0.9519   \\ \hline
		\bf User 6  &0.9741 &0.7741& 0.9397& 0.9102& 0.9306 &0.5959& 0.8894& 0.8000& 0.6979&  0.9636& 0.9682 &0.7955& 0.9829 &0.9846 &0.9282 \\ \hline \hline
		\multicolumn{16}{c}{\bf Naive Bayes Classifier} \\ \hline \hline
		\bf User 1  &0.9019 &0.6981  &0.9000& 0.8049 &0.6146& 0.4732& 0.6800& 0.4467 &0.3467 &0.5000 &0.6474& 0.5421 & 0.8789& 0.7046& 0.8844    \\ \hline
		\bf User 2  &0.9099 &0.7171  &0.9027  &0.7087 &0.5087 &0.5000 &0.8174& 0.6478& 0.7391& 0.8412& 0.8412& 0.6118& 0.9109& 0.9036& 0.8982   \\ \hline
		\bf User 3  &0.9518 &0.7571 &0.6589  &0.9250 &0.7300 &0.6950 &0.8889 &0.6778 &0.7500 &0.7846 &0.5949 &0.7897& 0.8912& 0.6421& 0.8930    \\ \hline
		\bf User 4  &0.9279  &0.7423 &0.9261& 0.9429 &0.7381& 0.4952 &0.8750 &0.6600 &0.6050 &0.7500  &0.7312 &0.7250& 0.9296  &0.7315 &0.7278
		\\ \hline
		\bf User 5  &0.9219& 0.7200&  0.9105 &0.6176 &0.6176& 0.6294& 0.8903& 0.6710& 0.7355& 0.6194& 0.8194& 0.6258& 0.9308 &0.7231& 0.9269   \\ \hline
		\bf User 6  & 0.8741 &0.6862& 0.8931& 0.6490& 0.4735& 0.4776& 0.7319& 0.4809& 0.6723& 0.1955& 0.3955& 0.6273& 0.6735  &0.6803 &0.8906  \\ \hline \hline
		\multicolumn{16}{c}{\bf KNN Classifier} \\ \hline \hline
		\bf User 1  &0.7333& 0.7389& 0.6852& 0.7122& 0.4732& 0.2585& 0.5800& 0.3467& 0.3667& 0.8316 &0.8211& 0.6895& 0.9450 &0.9541 &0.9376   \\ \hline
		\bf User 2  &0.9604& 0.7477 &0.9045& 0.5217& 0.2783 &0.1826 &0.8391& 0.6087 &0.4435 &0.8059 &0.8412 &0.8059& 0.9691 &0.9691 &0.9164   \\ \hline
		\bf User 3  &0.9643 &0.7571  &0.9054 &0.5400 &0.5900 &0.2050& 0.7222 &0.5222 &0.3389 &0.8051 &0.6103 &0.7436& 0.9561 &0.9491 &0.9316   \\ \hline
		\bf User 4  &0.9568 &0.9658 &0.9351& 0.8333& 0.6333& 0.1333& 0.4450& 0.6250 &0.6350& 0.6813 &0.4813& 0.7688& 0.8741 &0.6889 &0.8481   \\ \hline
		\bf User 5  &0.9638 &0.7600 & 0.9219 &0.6765& 0.4118& 0.3000& 0.6452 &0.6581 &0.5032 &0.6387 &0.6645& 0.7355& 0.9769 &0.7769& 0.9558   \\ \hline
		\bf User 6  &0.9397 &0.7466 &0.8759& 0.5673& 0.3143& 0.2408 &0.5064& 0.4340& 0.3234 &0.8773 &0.8773& 0.5318& 0.7316& 0.9350 &0.9009   \\ \hline \hline
		\multicolumn{16}{c}{\bf SVM Classifier} \\ \hline \hline
		\bf User 1  &0.7481& 0.7370& 0.9315& 0.8049& 0.6732& 0.6000& 0.8400& 0.6200& 0.7333& 0.7263& 0.9158& 0.7474& 0.9688& 0.9761 &0.9431  \\ \hline
		\bf User 2  &0.9622 &0.7658& 0.9279 &0.6522 &0.6609 &0.8261 &0.9435& 0.7565 & 0.7696 &0.8824& 0.6882 &0.8059& 0.9855&  0.9855& 0.9327  \\ \hline
		\bf User 3  &0.9714  &0.7679 & 0.9089 &0.7500 &0.7250 &0.7450 &0.8556 & 0.6333& 0.6889& 0.8974& 0.6769& 0.7538& 0.9719 &0.9579 &0.9246  \\ \hline
		\bf User 4  &0.9838& 0.7892&  0.9640& 0.9238 &0.7286& 0.3095& 0.9350& 0.7150 &0.7500& 0.8500& 0.8375& 0.5938& 0.9685 &0.9685 &0.9537  \\ \hline
		\bf User 5  &0.9581  &0.7562& 0.9067 &0.9353& 0.7235& 0.5000& 0.9226& 0.6968& 0.7484 &0.8968 &0.9161 &0.8194& 0.9673 &0.7769 &0.9596     \\ \hline
		\bf User 6  &0.9552 &0.7638  &0.9414 &0.6939 &0.9020 &0.4082& 0.8851 &0.6638 &0.4511 &0.9636 &0.9364 &0.7045& 0.9573 &0.9607& 0.9162 \\ \hline \hline
		\multicolumn{16}{c}{\bf Ensemble Learning-BAG Classifier} \\ \hline \hline
		\bf User 1  &0.9870 &0.7852  &0.9481 &0.7122 &0.7317 &0.7707 &0.9133 &0.7067& 0.5933 &0.9632 &0.9421 &0.7842& 0.9872 &0.7963 &0.9486  \\ \hline
		\bf User 2  &0.9838  &0.9874& 0.9604& 0.8870& 0.7261& 0.5652 &0.9783 &0.7826 &0.8174 &0.9588 &0.9647 &0.8706& 0.9873 &0.9909& 0.9491   \\ \hline
		\bf User 3  & 0.9786& 0.7732 &0.9179 &0.9450 &0.9500 &0.6400 &0.9278& 0.7111& 0.5556& 0.9077& 0.9026 &0.8154& 0.9807 & 0.9561& 0.9105  \\ \hline
		\bf User 4  &0.9874 & 0.9964 &0.9495 &0.9619 &0.7714 &0.6762 &0.9650& 0.7400& 0.8350& 0.9375& 0.9563 &0.6000& 0.9778 &0.9778& 0.9444   \\ \hline
		\bf User 5  &0.9867 &0.9886  &0.7276& 0.9059 &0.7235 &0.7118 &0.7290& 0.7161 &0.7871 &0.9677 &0.9548 &0.6452& 0.9942 &0.7846 &0.9462   \\ \hline
		\bf User 6  &0.9741  &0.7724 &0.7138 &0.6980 &0.9224& 0.5918 &0.8553 &0.8255& 0.7064 &0.9727& 0.9591 &0.5955& 0.9846 &0.9897 &0.9197   \\ \hline \hline
	\end{tabular}}
\end{table}	

Looking at Table I-IV, KNN, Decision trees and SVM perform much better in the personal model for both number of features with normalization and for overall accuracy prospect; decision trees outperformed then KNN and SVM. The highest obtained accuracies are in boldface. For further analysis the best-recommended system settings are 70 features normalized and randomly permuted used by Decision tree and SVM classifier to recognize the individual's activity on a smartwatch. Given a fixed amount of data, the performance of KNN and SVM decreases significantly when the window size is increased as it reduces the number of training samples but this does not affect decision tree classifier. Since decision tree, KNN and SVM classifiers trained using 70 normalized and randomly permuted features happens to be the best setting for smartwatch based PAR system. 

Results are summarized in Fig. 2 which presents the overall classification accuracy with 98\% confidence interval. According to these results, decision tree classifier shows higher accuracy for every size of the window. The performance of SVM and KNN through good but is relatively less than that of the decision tree in all cases. Overall, decision trees performance happens to be best followed by SVM as their performance did not much degraded when changing the size of samples per window. On the other hand, the performance of KNN reduced significantly as the window size is increased. We think that it is because we have a fixed amount of data. So a bigger window size results in a smaller number of training samples. To confirm this, we repeated the same experiment for each activity individually. The entire results are summarized in Table I-IV which confirms our hypothesis. The additional experiments on single subject cross-validations are presented in Table V-VIII for both feature banks and personal and impersonal models respectively.

\begin{table*}[!ht]
	\tiny
	\caption{Single Subject cross-validation-based accuracy analysis for \textbf{Personal Model with 70 Features extracted using 75 samples per window}.}
\resizebox{\textwidth}{!}{\begin{tabular}{l c c c c c c c c c c c c c c c c}
		\hline \hline
		\multicolumn{1}{c}{\bf Users} & \multicolumn{3}{|c|}{\bf Walking} & \multicolumn{3}{c|}{\bf Walking Up-Stairs} & \multicolumn{3}{c|}{\bf Walking Down-Stairs} & \multicolumn{3}{c|}{\bf Running}& \multicolumn{3}{c}{\bf Jogging}\\ \hline
		& \multicolumn{1}{|c}{\bf NR-RP} & \multicolumn{1}{c}{\bf NR-NRP}& \multicolumn{1}{c|}{\bf UNR-RP} & \multicolumn{1}{c}{\bf NR-RP} & \multicolumn{1}{c}{\bf NR-NRP}& \multicolumn{1}{c|}{\bf UNR-RP}& \multicolumn{1}{c}{\bf NR-RP} & \multicolumn{1}{c}{\bf NR-NRP}& \multicolumn{1}{c|}{\bf UNR-RP}& \multicolumn{1}{c}{\bf NR-RP} & \multicolumn{1}{c}{\bf NR-NRP}& \multicolumn{1}{c|}{\bf UNR-RP}& \multicolumn{1}{c}{\bf NR-RP} & \multicolumn{1}{c}{\bf NR-NRP}& \multicolumn{1}{c}{\bf UNR-RP}\\ \hline  \hline
		\multicolumn{16}{c}{\bf Decision Trees Classifier} \\ \hline \hline
		\bf User 1  &0.9593& 0.9574 &0.7074 &0.6780& 0.6780& 0.7171& 0.5800& 0.5467& 0.7000 &0.6947 &0.9053 &0.7632& 0.9908 &0.9908 &0.9541  \\ \hline
		\bf User 2  &0.9748 &0.7640 &0.9405 &0.8913 &0.7217 &0.6130 &0.9217& 0.9522 &0.7739 &0.9235 &0.7235 &0.8176& 0.9909& 0.9873& 0.9218  \\ \hline
		\bf User 3  &0.9821 &0.7821  &0.9357 &0.9550 &0.7500 &0.4950 &0.9389 &0.7111 &0.7167& 0.8769 &0.8872& 0.7487& 0.9421 &0.9246& 0.9105   \\ \hline
		\bf User 4  &0.9820 &0.9856 &0.9586 &0.9286 &0.7143 &0.6762 &0.8200 &0.5950 &0.8100 &0.8562 &0.9000  &0.6062& 0.9778 &0.7741& 0.9389  \\ \hline
		\bf User 5  &0.9790& 0.7733 &0.9257& 0.8412& 0.6235& 0.8294 &0.6903& 0.8710& 0.7548& 0.9290& 0.9419 &0.6129& 0.9750 &0.9731 &0.9462   \\ \hline
		\bf User 6  & 0.9741& 0.9810 &0.9172 &0.7265 &0.7102 &0.3673 &0.7830 &0.7489& 0.6894& 0.8727 & 0.8773 &0.7773& 0.9641 &0.9675 &0.9248\\ \hline \hline
		\multicolumn{16}{c}{\bf Naive Bayes Classifier} \\ \hline \hline
		\bf User 1  &0.8593& 0.6333& 0.7870 &0.5073 &0.4634 &0.6537 &0.5467& 0.3733& 0.6200& 0.5368 &0.5632& 0.7211& 0.8789 &0.6642 &0.8917   \\ \hline
		\bf User 2  &0.8757 &0.8793 & 0.6505 &0.3609 &0.3826 &0.5696& 0.7478 &0.5739& 0.7739 &0.7176 &0.5529 &0.7941& 0.8636& 0.6727& 0.7000  \\ \hline
		\bf User 3  &0.9161& 0.7179 &0.6071 &0.8750& 0.6650 &0.6300 &0.7889& 0.5722 &0.4667 &0.5333 &0.6462& 0.7590& 0.8561& 0.6614 &0.8842  \\ \hline
		\bf User 4  &0.8955 &0.6991 &0.8541& 0.8381 &0.6524& 0.6000& 0.5050 &0.4550 &0.5400 &0.6000& 0.4063 &0.4625& 0.9130& 0.7148& 0.7333  \\ \hline
		\bf User 5  &0.8857 &0.8800 &0.8781 &0.5765 &0.3647& 0.2588& 0.8065& 0.5613& 0.6065& 0.7613 &0.5871 &0.7677& 0.9096& 0.7135 &0.9154 \\ \hline
		\bf User 6  &0.8638 &0.6879 &0.8448& 0.7224& 0.5429 &0.4612 &0.3362 &0.3064& 0.5745& 0.3273 &0.5318& 0.3955 & 0.8598& 0.6547 &0.8838
		\\ \hline \hline
		\multicolumn{16}{c}{\bf KNN Classifier} \\ \hline \hline
		\bf User 1  &0.9648 &0.9426 &0.8944 &0.4390& 0.3951 &0.1268 &0.4267& 0.4667& 0.1600& 0.7053 &0.7158& 0.1842& 0.9835 &0.9725 &0.9138\\ \hline
		\bf User 2  &0.9532 &0.9532& 0.8685& 0.6565& 0.5087& 0.1391& 0.5174& 0.5174 &0.2217 &0.5118 &0.7118& 0.2000& 0.9545 & 0.9600 &0.8927  \\ \hline
		\bf User 3  &0.9625 &0.9732& 0.8536& 0.7700 &0.5950 &0.4650 &0.7389 &0.5278& 0.1444& 0.4718& 0.5795& 0.2974 &0.9263& 0.9070 &0.8526   \\ \hline
		\bf User 4  &0.9550 &0.9568  &0.7964& 0.5048 &0.4810& 0.2048& 0.3450 &0.5200 &0.2700 &0.3250 &0.2562& 0.0938& 0.9481& 0.9574& 0.9204   \\ \hline
		\bf User 5  & 0.9771 &0.7638  &0.7048 &0.5471 &0.3176& 0.1353& 0.7161 &0.5484 &0.1613 &0.7742& 0.7935& 0.5935 &0.9635 &0.9654 &0.9058  \\ \hline
		\bf User 6  &0.9603 &0.9759  &0.6190 &0.6898 &0.4612 &0.0980 &0.2213 &0.2043 &0.0596 &0.6955& 0.4727& 0.2182 &0.9675& 0.9556& 0.8940   \\ \hline \hline
		\multicolumn{16}{c}{\bf SVM Classifier} \\ \hline \hline
		\bf User 1  &0.9778 &0.7759& 0.8852& 0.8634 &0.7317 &0.4000 &0.8067 &0.8067 &0.5600 &0.9000& 0.9263& 0.5105 & 0.9706 &0.9725 &0.9193  \\ \hline
		\bf User 2  &0.9946 &0.9910 & 0.9297 &0.9522 &0.7652 &0.4826 &0.9348 &0.7435 &0.4870 &0.9000 &0.7294  &0.5353& 0.9909& 0.9855& 0.9200  \\ \hline
		\bf User 3  &0.9804 &0.9839 &0.8732 &0.9150 &0.7200& 0.6150& 0.9333 &0.9389& 0.4333& 0.8718 &0.8769 &0.7231& 0.9439 &0.7175 &0.9123 \\ \hline
		\bf User 4  &0.9820 &0.9748 &0.9063 &0.9667 &0.7619 &0.6048 &0.9100& 0.7000& 0.7100& 0.9313& 0.9437 &0.7375& 0.9667 &0.7778& 0.9241  \\ \hline
		\bf User 5  &0.9829 &0.9752 &0.6990& 0.6882& 0.6588& 0.4765& 0.8968& 0.7032 &0.4516 &0.9097 &0.6968& 0.7677& 0.9827 &0.9808 &0.9346     \\ \hline
		\bf User 6  &0.9759 &0.7862 & 0.8983 &0.9510 &0.9633 &0.5673 &0.6851& 0.6638 &0.5872& 0.9273& 0.9545 &0.7091& 0.9675 &0.9641 &0.8872 \\ \hline \hline
		\multicolumn{16}{c}{\bf Ensemble Learning-BAG Classifier} \\ \hline \hline
		\bf User 1  &0.9556 &0.9500 &0.9241 &0.8780 &0.6829 &0.7317 &0.7333 &0.5133 &0.6467& 0.8947 &0.8842 &0.8000& 0.9908 &0.9890& 0.9596  \\ \hline
		\bf User 2  &0.9658 &0.7730& 0.9405& 0.8826& 0.7087& 0.8261& 0.9087 &0.7174 &0.5696 &0.9353& 0.7529 &0.8000& 0.9873& 0.9873& 0.9364   \\ \hline
		\bf User 3  &0.9821 &0.7804 &0.9268 &0.9500& 0.7400& 0.4700& 0.8944& 0.7056& 0.5500 &0.8667 &0.8564 &0.7487 & 0.9386&  0.7228& 0.9211  \\ \hline
		\bf User 4  &0.9874& 0.9802 &0.9532 &0.6905 &0.6952 &0.6333 &0.8250 &0.5150 &0.6650 &0.8625 &0.9000 &0.6000& 0.9741 & 0.7741& 0.9444   \\ \hline
		\bf User 5  & 0.9619 &0.7695 &0.9390& 0.8235& 0.6000 & 0.7941& 0.8645& 0.6903& 0.7613 &0.9032 &0.9226 &0.8065 & 0.9750 &0.7750& 0.9519 \\ \hline
		\bf User 6  &0.9741 &0.9724 &0.7241 &0.9061 &0.6939 &0.6367 &0.8170& 0.7872& 0.6936& 0.9500 &0.8409 &0.5864& 0.9761 &0.9658 &0.7231   \\ \hline \hline
	\end{tabular}}
\end{table*}

\begin{table*}[!ht]
	\tiny
	\caption{Single Subject cross-validation-based accuracy analysis for \textbf{Impersonal Model with 70 Features extracted using 75 samples per window}.}
	\resizebox{\textwidth}{!}{\begin{tabular}{l c c c c c c c c c c c c c c c c}
		\hline \hline
		\multicolumn{1}{c}{\bf Users} & \multicolumn{3}{|c|}{\bf Walking} & \multicolumn{3}{c|}{\bf Walking Up-Stairs} & \multicolumn{3}{c|}{\bf Walking Down-Stairs} & \multicolumn{3}{c|}{\bf Running}& \multicolumn{3}{c}{\bf Jogging}\\ \hline
		& \multicolumn{1}{|c}{\bf NR-RP} & \multicolumn{1}{c}{\bf NR-NRP}& \multicolumn{1}{c|}{\bf UNR-RP} & \multicolumn{1}{c}{\bf NR-RP} & \multicolumn{1}{c}{\bf NR-NRP}& \multicolumn{1}{c|}{\bf UNR-RP}& \multicolumn{1}{c}{\bf NR-RP} & \multicolumn{1}{c}{\bf NR-NRP}& \multicolumn{1}{c|}{\bf UNR-RP}& \multicolumn{1}{c}{\bf NR-RP} & \multicolumn{1}{c}{\bf NR-NRP}& \multicolumn{1}{c|}{\bf UNR-RP}& \multicolumn{1}{c}{\bf NR-RP} & \multicolumn{1}{c}{\bf NR-NRP}& \multicolumn{1}{c}{\bf UNR-RP}\\ \hline  \hline
		\multicolumn{16}{c}{\bf Decision Trees Classifier} \\ \hline \hline
		\bf User 1  &0.7455 &0.7412 &0.3338 &0.9296 &0.9286 &0.3888 &0.7896 &0.7896 &0.7383 &0.3563 &0.3677 &0.3885& 0.8648 &0.8739 &0.9086   \\ \hline
		\bf User 2  &0.9370 &0.9458 &0.1308 &0.9682 &0.9665 &0.8129 &0.7210 &0.7223 &0.6605 &0.8600& 0.8524 &0.7094& 0.8445& 0.8320& 0.8367  \\ \hline
		\bf User 3  &0.9846 &0.9832 &0.5609 &0.7505 &0.7471 &0.6240 &0.6268 &0.6311& 0.5557& 0.8437& 0.8508& 0.7061& 0.7346& 0.7392 &0.6330   \\ \hline
		\bf User 4  &0.9524 &0.9532 & 0.6688 &0.7906 &0.7850 &0.4028 &0.6485 &0.6470 &0.3876 &0.2348& 0.2274 &0.1835& 0.9762 &0.9773 &0.9486  \\ \hline
		\bf User 5  &0.9617& 0.9636&  0.8874& 0.8023& 0.7849 &0.1552 &0.6791& 0.6747& 0.5766 &0.7264& 0.7157 &0.7535& 0.9442& 0.9456 &0.9133   \\ \hline
		\bf User 6  &0.8523& 0.8491 &0.4219 &0.6711 &0.6643  &0.3040& 0.3448 & 0.3552 &0.5130 &0.7254 &0.7513 &0.4955& 0.7814& 0.7879& 0.8374 \\ \hline \hline
		\multicolumn{16}{c}{\bf Naive Bayes Classifier} \\ \hline \hline
		\bf User 1  &0.6192 &0.8429& 0.4455& 0.7087& 0.1408& 0.3883& 0.5649& 0.1039& 0.5649& 0.2396 &0.1719 &0.3646& 0.7646 &0.8102& 0.8266   \\ \hline
		\bf User 2  &0.9123& 0.7728& 0.3953 &0.5536 &0.1330 &0.5880 &0.6609 &0.1116 &0.5536 &0.5118 &0.1294 &0.7118& 0.8582 &0.8873& 0.8273  \\ \hline
		\bf User 3  &0.8286 &0.6982 &0.5643 &0.4216 &0.1961 &0.5392 &0.5683 &0.2022 &0.4481 &0.5736 &0.2792& 0.6954& 0.7902 &0.7115 &0.7413  \\ \hline
		\bf User 4  &0.8342 &0.5802 &0.4432 &0.5164 &0.0845 &0.5164& 0.5149 &0.1535& 0.3960& 0.3415 &0.0183 &0.2378 & 0.8965 &0.9279 &0.9205 \\ \hline
		\bf User 5  &0.6724 &0.8114 &0.1410& 0.5291& 0.1279& 0.5756 &0.6076 &0.1266 &0.4873& 0.4969 &0.1698 &0.7610& 0.7404& 0.8769& 0.6308 \\ \hline
		\bf User 6  &0.7126& 0.6713 &0.3270 &0.4859 &0.1767 &0.3735 &0.2594 &0.0879& 0.3347& 0.3527 &0.2188 &0.4955& 0.8222 &0.5573& 0.8393  \\ \hline \hline
		\multicolumn{16}{c}{\bf KNN Classifier} \\ \hline \hline
		\bf User 1  &0.8096 &0.8096& 0.8429& 0.6408 &0.6408& 0.1408& 0.4026& 0.4026& 0.1039& 0.2448& 0.2448 &0.1719& 0.8029&  0.8029 &0.8102\\ \hline
		\bf User 2  &0.9141 &0.9141& 0.7728& 0.6438& 0.6438& 0.1330& 0.5794 &0.5794 &0.1116 &0.3235 &0.3235&  0.1294& 0.9182 &0.9182& 0.8873  \\ \hline
		\bf User 3  &0.9143 &0.9143& 0.6982& 0.5931 &0.5931 &0.1961 &0.5683 &0.5683& 0.2022 &0.5381 &0.5381 &0.2792& 0.8759 &0.8759& 0.7115   \\ \hline
		\bf User 4  & 0.9387 &0.9387  &0.5802& 0.4131 & 0.4131 &0.0845& 0.2574 &0.2574 &0.1535 &0.1037 &0.1037 &0.0183& 0.9593& 0.9593& 0.9279  \\ \hline
		\bf User 5  &0.9733 &0.9733 &0.8114 &0.4767 &0.4767 &0.1279 &0.5380& 0.5380 &0.1266 &0.3019 &0.3019 &0.1698& 0.9231 &0.9231 &0.8769   \\ \hline
		\bf User 6  &0.8176 &0.8176 &0.6713& 0.4618 &0.4618 &0.1767& 0.2008 &0.2008 &0.0879 &0.3705 &0.3705 &0.2188& 0.7726& 0.7726 &0.5573   \\ \hline \hline
		\multicolumn{16}{c}{\bf SVM Classifier} \\ \hline \hline
		\bf User 1  &0.9372 &0.9372& 0.6007 &0.9417& 0.9417 &0.2233 &0.7857 &0.7857& 0.5909& 0.6146& 0.6146& 0.5313& 0.9197 &0.9197 &0.8832  \\ \hline
		\bf User 2  &0.9517 &0.9517& 0.9660& 0.9828 &0.9828& 0.0258 &0.8712 &0.8712 & 0.6266& 0.9059 &0.9059 &0.6882& 0.9655  &0.9655 &0.8909  \\ \hline
		\bf User 3  & 0.9875 &0.9875 &0.8750& 0.8578& 0.8578& 0.1716 &0.8852  &0.8852 &0.5355& 0.9137& 0.9137& 0.6802& 0.8776 &0.8776 &0.7483
		\\ \hline
		\bf User 4  &0.9928 &0.9928 & 0.6883& 0.8685 &0.8685 &0.2066 &0.8861 &0.8861& 0.4208& 0.8049 &0.8049 &0.5183& 0.9778 &0.9778 &0.9482  \\ \hline
		\bf User 5  &0.9867 &0.9867& 0.9543& 0.8663& 0.8663 &0.0291 &0.8291& 0.8291  &0.5886 &0.7358&  0.7358 &0.5849& 0.9654& 0.9654& 0.9000     \\ \hline
		\bf User 6  &0.9639 &0.9639 &0.9466& 0.8996 &0.8996 &0.0723 &0.6820 &0.6820 &0.2176 &0.8527 &0.8527 &0.5670& 0.9573& 0.9573& 0.8256 \\ \hline \hline
		\multicolumn{16}{c}{\bf Ensemble Learning-BAG Classifier} \\ \hline \hline
		\bf User 1  &0.7423 &0.7240& 0.3532 &0.9194 &0.9301& 0.3888 &0.7721& 0.7825& 0.7123& 0.3219& 0.3255& 0.3906& 0.8746 &0.8611 &0.9111  \\ \hline
		\bf User 2  & 0.9331 &0.9392 &0.1644 &0.9665& 0.9597& 0.7931& 0.7120 &0.7206 &0.6652 &0.8288  &0.8253 &0.7141& 0.8393  &0.8480 &0.8349  \\ \hline
		\bf User 3  &0.9832& 0.9852& 0.5695& 0.7333& 0.7216 &0.6039& 0.6186& 0.6235& 0.5322& 0.8376& 0.8365&  0.7015& 0.7362 &0.7392 &0.6451   \\ \hline
		\bf User 4  &0.9450 &0.9505& 0.6701 &0.7690 &0.7643 &0.3911 &0.6416 &0.6391 &0.3856 &0.2305 &0.2329 &0.1951& 0.9758& 0.9747& 0.9444    \\ \hline
		\bf User 5  &0.9598 &0.9577 &0.8851& 0.7628& 0.7756 &0.1651 &0.6918 &0.6842 &0.5766 &0.7182 &0.7170 &0.7440& 0.9415 &0.9463 &0.9137  \\ \hline
		\bf User 6  &0.8522 &0.8515& 0.4389& 0.6627 &0.6514 &0.2892& 0.3477 &0.3485 &0.5188 &0.7201 &0.7076 &0.4906& 0.7805 & 0.7701 &0.8487   \\ \hline \hline
	\end{tabular}}
\end{table*}

\section{Conclusion}
\label{sec8}

Feature banks created by different smartphone-based physical activity recognition studies can be used for physical activity recognition on smartwatches because smartphone also uses accelerometer, gyroscope and magnetometer sensors and hence the same set of features is acceptable for a smartwatch. Decision trees, SVM and KNN showed the best results with a minor difference. Additionally, Naive Bayes and ensemble learning with Bag classifiers also produce a good performance. Furthermore, feature normalization and random permutation processes significantly improve the classification performance in several cases. In addition to above, without any doubt, changing window size affects the recognition accuracy.

From results, we observed that every activity and each classifier's results of recognition is highly correlated with window size. In general, each classifier performs well when features are extracted using 25-125 samples per window. Personal models perform better for smartwatch based physical activity recognition providing an average accuracy of 98\%. Furthermore, the impersonal model produces 94\% and 90\% accuracy respectively with the same settings of the personal model. Finally, the recommended combination for smartwatch based physical activity recognition system is decision tree classifier trained using 25-125 samples per window of normalized and randomly permuted features. In our future work, we will investigate the effects of several linear/non-linear supervised and unsupervised feature selection methods.

\end{document}